\documentclass[11pt]{article}

\usepackage[cp1251]{inputenc}

\begin{document}

\begin{center}

{\bf  E.M.  Ovsiyuk, V.M.  Red'kov \\[3mm]
SPHERICAL WAVES  OF SPIN  1 PARTICLE \\ IN ANTI DE SITTER SPACE-TIME
\\[3mm]
 Institute of Physics \\National
Academy of Sciences of Belarus\\
mozlena@tut.by,  redkov@dragon.bas-net.by
}
\end{center}

\begin{quotation}

Three  possible techniques to deal with a vector particle in the anti de Sitter cosmological model are viewed:
 Duffin --  Kemmer  --  Petiau matrix formalism    based on the  general tetrad recipe,
 group theory 5-dimensional approach based on the symmetry group SO(3.2), and
a  tetrad form of Maxwell equations
  in complex  Riemann -- Silberstein -- Majorana -- Oppenheimer representation.
In the first part,  a spin 1 massive field  is considered in static coordinates of the anti de Sitter space-time
in  tetrad-based  approach.
The complete set of  spherical wave solutions with quantum numbers $(\epsilon , j,m,l)$
 is constructed; angular dependence in wave functions is described with the help of
Wigner functions.  The energy quantization rule has been found.
Transition to massless case of electromagnetic field is specified,
 and electromagnetic solutions in Lorentz gauge have been constructed.
In the second part, the  problem of a particle with  spin
 1 is considered on the base of  5-dimensional   wave equation
  specified in the same static coordinates.
  In the third part,  a rarely used   approach, based on tetrad form of Maxwell equations
  in complex    representation is examined in the anti de Sitter model.

\end{quotation}


  SO(3.2) group, 5-dimensional equation,   Majorana -- Oppenheimer representation,

\vspace{5mm}


\section{Introduction }

Examining  fundamental  particle fields on the background of expanding universe,
in particular de Sitter and anti de Sitter models,
has a long history; special value of these  geometries consists in their simplicity and high symmetry
groups underlying them
which makes us to believe in existence of exact analytical treatment for  some fundamental problems of
classical and quantum field theory  in these curved spaces. In particular, there exist special representations for
fundamental wave equations, Dirac's and Maxwell's, which are explicitly invariant under symmetry groups
$SO(4.1)$  and $SO(3.2)$ for these models.
In the most of the literature, when dealing with a spin 1 field  in de Sitter models
they use group theory approach.
Many of the  most important references are  given below
(it is not  exhaustive bibliography, which  should be enormous):

Dirac \cite{Dirac-1935, Dirac-1936},
Schr\"{o}dinger \cite{Schrodinger-1939, Schrodinger-1940},
Lubanski--Rosenfeld \cite{Lubanski-Rosenfeld-1942},
Goto \cite{Goto-1951},
Ikeda \cite{Ikeda-1953},
Nachtmann \cite{Nachtmann},
Chernikov--Tagirov \cite{Chernikov-Tagirov},
Geheniau--Schomblond\cite{Geheniau-Schomblond},
Bor\-ner--Durr \cite{Borner-Durr-1969},
Tugov   \cite{Tugov-1969},
Fushchych--Krivsky  \cite{Fushchych-Krivsky-1969},
Chevalier   \cite{Chevalier-1970},
Castag\-nino  \cite{Castagnino-1970, Castagnino-1972},
Vidal  \cite{Vidal-1970},
Adler \cite{Adler-1972},
Schnirman--Oliveira  \cite{Schnirman-Oliveira-1972},
Tagirov   \cite{Tagirov-1973},
Riordan   \cite{Riordan-1974},
Pestov--Chernikov--Shavoxina  \cite{Pestov-Chernikov-Shavoxina-1975},
Candelas--Raine   \cite{Candelas-Raine-1975},
Schom\-blond--Spindel    \cite{Schomblond-Spindel-1976, Schomblond-Spindel-1976'},
Dowker--Critchley    \cite{Dowker-Critchley-1976},
Avis-Isham--Storey   \cite{Avis-Isham-Storey-1978},
Brugarino  \cite{Brugarino-1980},
Fang-Fronsdal   \cite{Fang-Fronsdal-1980},
Angelopoulos et al  \cite{Angelopoulos-Flato-Fronsdal-Sternheimer-1981},
Burges   \cite{Burges-1984},
Deser-Nepo\-mechie   \cite{Deser-Nepomechie-1984},
Dullemond-Beveren \cite{Dullemond-Beveren-1985},
Gazeau  \cite{Gazeau-1985},
Allen    \cite{Allen-1985},
Fef\-ferman--Graham   \cite{Fefferman-Graham-1985},
Flato--Fronsdal--Gazeau    \cite{Flato-Fronsdal-Gazeau-1986},
Allen--Jacobson     \cite{Allen-Jacobson-1986},
Al\-len--Folac\-ci   \cite{Allen-Folacci-1987},
Sanchez    \cite{Sanchez-1987},
Pathinayake--Vilenkin--Allen   \cite{Pathinayake-Vilenkin-Allen-1988},
Gazeau--Hans    \cite{Gazeau-Hans-1988},
Bros--Gazeau--Moschella   \cite{Bros-Gazeau-Moschella-1994},
Takook  \cite{Takook-1997},
Pol'shin  \cite{Pol'shin-1998(1), Pol'shin-1998(2), Pol'shin-1998(3)},
Ga\-zeau--Takook  \cite{Gazeau-Takook-2000},
Takook   \cite{Takook-2000},
Deser--Waldron   \cite{Deser-Waldron-2001, Deser-Waldron-2001'},
Spradlin--Strominger--Volovich  \cite{Spradlin-Strominger-Volovich-2001},
Cai--Myung--Zhang    \cite{Cai-Myung-Zhang-2002},
Garidi--Huguet--Renaud   \cite{Garidi-Huguet-Renaud-2003},
Rouhani--Takook  \cite{Rouhani-Takook-2005},
Behroozi et al  \cite{Behroozi-Rouhani-Takook-Tanhayi-2006},
Huguet--Queva--Renaud   \cite{Huguet-Queva-Renaud-2006},
 Garidi et al   \cite{Garidi-Gazeau-Rouhani-Takook-2008},
Huguet--Queva--Renaud   \cite{Huguet-Queva-Renaud-2008(2)},
Dehghani et al  \cite{Dehghani-Rouhani-Takook-Tanhayi-2008},
Moradi-Rouhani-Takook   \cite{Moradi-Rouhani-Takook-2008},
Faci et al \cite{Faci-Huguet-Queva-Renaud-2009}

The interest to exact solutions of wave equations for particles with different  spins in de Sitter  space was greatly
increased in connection with  Hawking radiation
 \cite{Hawking-1971, Hawking-1974, Hawking-1975}.
 De Sitter cosmological model admits     exact treatment in contrast to black hole space-time geometry:

Lohiya--Panchapakesan \cite{Lohiya-Panchapakesan-1978, Lohiya-Panchapakesan-1979},
 Khanal--Panchapakesan \cite{Khanal-Panchapakesan-1981(1), Khanal-Panchapakesan-1981(2)},
  Khanal \cite{Khanal-1983, Khanal-1985},
   Hawking--Page \cite{Hawking-Page-1983},
   Chandrasekhar \cite{Chandrasekhar-1983},
    Otchik \cite{Otchik-1985}, Motolla
\cite{Motolla-1985}, Bogush--Otchik--Red'kov \cite{Bogush-Otchik-Red'kov-1986},
Mishima--Nakayama \cite{Mishima-Nakayama-1987},
Polarski \cite{Polarski-1989},
Suzuki--Takasugi \cite{Suzuki-Takasugi-1996},
 Suzuki--Taka\-sugi--Umetsu \cite{Suzuki-Takasugi-Umetsu-1998, Suzuki-Takasugi-Umetsu-1999, Suzuki-Takasugi-Umetsu-2000}

The case of anti de Sitter space seems to be less examined though it is also very interesting
because of its topological properties. Any energy spectrum must be discreet,
 besides this geometry admits elliptical interpretation
 and physical  manifestation of that in cosmological context is also of great importance.
In the context of Hawking effect the most investigators used the Newman--Penrose formalism
\cite{Newman-1961},  \cite{Goldberg-Macfarlane-Newman-Rohrlich-Sudarshan-1967},
\cite{Chandrasekhar-1983}, \cite{Penrose-Rindler-1984}.

 Turning to the case of vector particle, we should note
 that  many years ago a matrix  Duffin -- Kemmer -- Petiau formalism was developed to treat a spin 1  field, it
 has a long and rich  history inseparably linked with  description
of  photons and  mesons:

De Broglie \cite{De Broglie-1934(1), De Broglie-1934(2)}, De Broglie--Winter
\cite{De Broglie-Winter-1934},
Petiau \cite{Petiau-1936},
Proca \cite{Proca-1938, Proca-1946},
Duffin \cite{Duffin-1938},
Kemmer \cite{Kemmer-1939, Kemmer-1943},
Bhabha \cite{Bhabha-1939},
Belinfante \cite{Belinfante-1939(1), Belinfante-1939(2)},
Sakata--Taketani \cite{Sakata-Taketani-1940},
Tonnelat \cite{Tonnelat-1941},
Schr\"{o}dinger \cite{Schrodinger-1943(1), Schrodinger-1943(2)},
Hitler \cite{Hitler-1943},
Harish-Chand\-ra \cite{Harish-Chandra-1946, Harish-Chandra-1947(1), Harish-Chandra-1947(2)},
Hoffmann \cite{Hoffmann-1947},
Utiyama \cite{Utiyama-1947},
Gel'fand--Yaglom \cite{Gel'fand-Yaglom-1948},
Schou\-ten \cite{Schouten-1949},
Gupta \cite{Gupta-1950},
Bleuler \cite{Bleuler-1950},
Fujiwara \cite{Fujiwara-1955},
Borgardt \cite{Borgardt-1956, Borgardt-1958},
Kuo\-hsien \cite{Kuohsien-1957},
Hjalmars \cite{Hjalmars-1961},
Bogush--Fedorov \cite{Bogush-Fedorov-1962},
Beckers--Pirotte \cite{Beckers-Pirotte-1968},
Casanova \cite{Casanova-1969},
Krivski--Romamenko--Fushchych \cite{Krivski-Romamenko-Fushchych-1969},
Goldman--Tsai--Yil\-diz \cite{Goldman-Tsai-Yildiz-1972},
Fushchych--Nikitin \cite{Fushchych-Nikitin-1977},

 However, this technique
till recent time was not  used in the curved space-time when constructing explicit solutions, though that possibility was
known -- see Weinberg \cite{Weinberg-1972},
Birrel-Davies \cite{Birrel-Davies-1982}.
 The situation is changing now:
Lunardi et al e \cite{Lunardi-Pimentel-Teixeira-Valverde-2000, Lunardi-Pimentel-Valverde-Manzoni-2002,
Casana-Pimente-Lunardi-Teixeira-2002, Casana-Fainberg-Pimentel-Lunardi-Teixeira-2003},
 Fainberg--Pimentel \cite{Fainberg-Pimentel-2000},
 De Montigny et ala
 \cite{De Montigny-Khanna-Santana-Santos-Vianna-2000},
    Red'kov \cite{Red'kov-1998(1), Red'kov-1998(2)},
   Bogush et al \cite{Bogush-Otchik-Red'kov-1986, Bogush-Kisel-Tokarevskaya-Red'kov-2002(1),
   Bogush-Kisel-Tokarevskaya-Red'kov-2002(2)}

In the present  paper, this approach  will be applied to
the case of spin 1 particle in anti de Sitter space; previously,
analogous treatment to a particle in de Sitter model was given  in
\cite{Bogush-Otchik-Red'kov-1986}.
Then we will treat the same problem on the base of 5-dimensional  wave equation.
And in the end we will consider the problem  within an approach, based on tetrad form of Maxwell equations
  in complex    representation \cite{Red'kov-Bogush-Tokarevskaya-Spix-2009, Bogush-Krylov-Ovsiyuk-Red'kov-2009},
  it seems be the most simple way to describe the classical  electromagnetic field in curved  space models.

\section{ Duffin -- Kemmer -- Petiau tetrad equation \\ in Riemannian space-time }

 We start from a flat
space equation  in  its  matrix  DKP-form
\begin{eqnarray}
(\; i\;  \beta ^{a} \; \partial_{a} \; - \; {m c \over  \hbar} \; )\;  \Phi  (x) =
0 \; ;
\label{2.1}
\end{eqnarray}

\noindent where
\begin{eqnarray}
\Phi  = ( \Phi _{0} , \; \Phi _{1} ,\; \Phi _{2}, \; \Phi _{3} ;
\; \Phi _{01}, \; \Phi _{02}, \; \Phi _{03}, \; \Phi _{23},\; \Phi
_{31}, \; \Phi _{12} )                     \;  ,
\nonumber
\\
\beta ^{a} = \left | \begin{array}{cc} 0 & \kappa ^{a} \\ \lambda
^{a} & 0
\end{array} \right | =
  \kappa  ^{a} \oplus \lambda  ^{a}  \;,
\;\; (\kappa  ^{a})_{j} ^{[mn]} \; = \; - i\; ( \delta ^{m}_{j} \;
g^{na} \;  - \; \delta
  ^{n}_{j} \; g^{ma} )\;\; ,
\nonumber
\\
( \lambda  ^{a})^{j}_{[mn]} \; = \;
  - i\; ( \delta  ^{a}_{m} \; \delta ^{j}_{n}  -
\delta ^{a}_{n} \;\delta ^{j}_{m} ) \; = - i\;\delta ^{aj}_{mn} \;
;
\label{2.2a}
\end{eqnarray}

\noindent $( g^{na} ) = \hbox{diag}( +1,-1,-1,-1 )$. The basic
properties of $\beta ^{a}$ are
\begin{eqnarray}
\beta ^{c}  \beta ^{a}  \beta ^{b} = \left| \begin{array}{cc}
     0 & \kappa ^{c} \lambda ^{a}\kappa ^{b} \\
\lambda^{c}  \kappa ^{a}  \lambda ^{b} & 0
\end{array} \right | ,  \;
(\lambda ^{c}  \kappa ^{a}  \lambda ^{b}) ^{j}_{[mn]} =
 i\; (  \delta ^{cb}_{mn}  g^{aj}   -  \delta ^{cj}_{mn}   g^{ab} ) \; ,
\nonumber
\\
(\kappa ^{c}  \lambda ^{a}  \kappa ^{b})^{[mn]}_{j}  =
 i  [ \delta ^{a}_{j} (g^{cm}  g^{bn}  -  g^{cn} g^{bm} )  +
g^{ac} ( \delta ^{n}_{j}  g^{mb}   -  \delta ^{m}_{j}
g^{nb} ) ] \;  ,
\label{2.2b}
\end{eqnarray}

\noindent and
\begin{eqnarray}
 \beta ^{c} \; \beta ^{a} \; \beta  ^{b} \;  + \;
\beta  ^{b} \; \beta ^{a} \; \beta ^{c}  =  \beta ^{c} \; g^{ab}
\; + \; \beta ^{b} g^{ac}\;  ,
\nonumber
\\
\; [\beta ^{c} , j^{ab} ] =  g^{ca} \;\beta^{b} \; - \; g^{cb} \;
\beta ^{a}\;   , \qquad j^{ab} =  \beta ^{a} \; \beta ^{b} \; - \;
\beta ^{b} \; \beta ^{a}\;  ,
\nonumber
\\
\;[j^{mn}, j^{ab}]  = ( \;g^{na} \; j^{mb} \;  - \; g^{nb} \; j^{ma}
\;) \; - \; (\; g^{ma} \;j^{nb}\;   - \; g^{mb} \; j^{na}\; )\;  .
\label{2.2c}
\end{eqnarray}

In accordance with tetrad recipe one  should generalize the
DKP-equation as follows \cite{Book-2009}
\begin{eqnarray}
[ \; i \; \beta ^{\alpha }(x)\; (\partial_{\alpha} \;  +  \;
B_{\alpha }(x) ) \; - {m c \over  \hbar} \;  ] \;\Phi  (x)  = 0 \; ,
\nonumber
\\
\beta ^{\alpha }(x) = \beta ^{a} e ^{\alpha }_{(a)}(x) \; , \;\;
B_{\alpha }(x) = {1 \over 2}\; j^{ab} e ^{\beta }_{(a)}\nabla
_{\alpha }( e_{(b)\beta })  \;  .
\label{2.3'}
\end{eqnarray}

\noindent This equation  contains  the  tetrad $e^{\alpha
}_{(a)}(x)$ explicitly. Therefore, there must  exist
a~possibility  to  demonstrate
 the~equivalence of  any  variants  of  this  equation
associated with various tetrads:
\begin{eqnarray}
e^{\alpha }_{(a)}(x)  \;\; , \qquad   e'^{\alpha }_{(b)}(x)\; = \;
 L^{\;\;b}_{a} (x) \; e^{\alpha }_{(b)}(x)  \; ,
\label{2.4a}
\end{eqnarray}

\noindent   $L^{\;\;b}_{a} (x)$  is a local Lorentz  transformation. We will  show  that  two such equations
\begin{eqnarray}
 [\;  i \beta
^{\alpha }(x) \; (\partial_{\alpha}  + B_{\alpha }(x)) \;  - \; {m c \over  \hbar}
\;] \; \Phi  (x)  = 0 \; ,
\nonumber
\\
\; [\; i \beta'^{\alpha }(x) \; (\partial_{\alpha}  + B'_{\alpha
}(x))  - {m c \over  \hbar} \; ]\;  \Phi'(x)  = 0 \; ,
\label{2.4b}
\end{eqnarray}

\noindent generated in tetrads $e^{\alpha }_{(a)}(x)$    and
 $e'^{\alpha }_{(b)}(x)$
respectively,  can  be  converted  into  each  other  through  a local gauge
transformation:
\begin{eqnarray}
\Phi '(x) = \left | \begin{array}{c}
                   \phi'_{a}(x) \\ \phi'_{[ab]}(x)
\end{array} \right | =
\left | \begin{array}{cc}
         L_{a}^{\;\;l} & 0 \\
         0 &  L_{a}^{\;\;m} L_{b}^{\;\;n}
\end{array} \right | \;\;
\left | \begin{array}{c}
                   \phi_{l}(x) \\ \phi_{[mn]}(x)
\end{array} \right |      \; .
\label{2.4c}
\end{eqnarray}

Indeed, tetrad DKP-equation  manifests  a~gauge  symmetry under local
Lorentz transformations in  complete analogy with more familiar
Dirac particle case. In  the~same time, the~wave function  from
this  equation  represents  scalar quantity relative to general
coordinate transformations:  if $\; x^{\alpha } \;
\rightarrow  \; x'^{\alpha } = f^{\alpha }(x)$ ,  then $\;
\Phi'(x) = \Phi (x)$.

It remains to demonstrate that this $DKP$  formulation  can  be
inverted into the Proca formalism in terms of  general  relativity
tensors. To this end, as a~first  step,  let  us  allow  for
the~sectional structure of $\beta ^{a}, J^{ab}$  and $\Phi (x)$ in
the~$DKP$-equation; then instead of (\ref{2.3'}) we get
\begin{eqnarray}
i \; [\; \lambda^{c} \; e^{\alpha }_{(c)} \; (\; \partial_{\alpha}
\; + \; \kappa ^{a} \; \lambda ^{b} \; e^{\beta }_{(a)} \;
\nabla_{\alpha }\; e_{(b)\beta }\; ) \;
]^{\;\;\;\;\;\;\;l}_{[mn]}\; \Phi _{l} = {m c \over  \hbar}\; \Phi _{[mn]}    \; ,
\nonumber
\\
i \; [\; \kappa ^{c} \; e^{\alpha }_{(c)} \; ( \;\partial_{\alpha} \;
+ \; \lambda ^{a} \; \kappa ^{b}\; e^{\beta }_{(a)} \; \nabla
_{\alpha } \; e_{(b)\beta }\; ) \; ]^{\;\;\;[mn]}_{l} \;
\Phi_{[mn]}  = {m c \over  \hbar} \Phi _{l} \; ,
\label{2.6a}
\end{eqnarray}

\noindent which lead to
\begin{eqnarray}
 (e_{(a)}^{\alpha} \; \partial_{\alpha} \; \Phi _{b} \; - \;
e^{\alpha }_{(b)} \; \partial_{\alpha}\; \Phi _{a}) \; + \; (
\gamma ^{c}_{\;\;ab} - \gamma ^{c}_{\;\;ba} ) \; \Phi _{c} = {m c \over  \hbar} \;
\Phi _{ab}\; ,
\nonumber
\\
e^{(b)\alpha } \;\partial_{\alpha } \; \Phi _{ab} \; + \; \gamma
^{nb} _{\;\;\;\;n} \Phi _{ab} + \gamma ^{\;\;mn}_{a} \Phi _{mn}  =
{m c \over  \hbar} \; \Phi _{a}  \; ;
\label{2.6b'}
\end{eqnarray}

\noindent the  symbol $\gamma _{abc}(x)$ is used to designate
Ricci  coefficients:
\begin{eqnarray}
 \gamma _{abc}(x) \; =  \; - \; e_{(a)\alpha
; \beta  } \; e^{\alpha }_{(b)} \; e^{\beta }_{(c)} \; .
\nonumber
\end{eqnarray}
 In
turn,  (\ref{2.6b'}) will look  as   the~Proca equations
\begin{eqnarray}
\nabla _{\alpha }\; \Psi _{\beta} - \nabla _{\beta }\;
\Psi_{\alpha } = m \; \Psi _{\alpha \beta } , \qquad \nabla
^{\beta }\; \Psi _{\alpha \beta } = m \; \Psi _{\alpha } \; ;
\label{Proca}
\end{eqnarray}

\noindent
they are rewritten in terms of tetrad  components
\begin{eqnarray}
\Phi _{a}  =  e^{\alpha }_{(a)}   \Phi _{\alpha }\; ,
\qquad \Phi _{ab}= e^{\alpha }_{(a)}    e^{\beta
}_{(b)} \;  \Phi_{\alpha \beta }     \; .
\label{2.7}
\end{eqnarray}

So, as evidenced by the~above, the~manner of introducing
the~interaction between a~spin  $1$  particle  and  external
classical gravitational field can be successfully unified
with~the approach  that occurred with regard to a~spin $1/2$
particle and was first  developed by  Tetrode,  Weyl,  Fock,
Ivanenko.  One  should   attach   great significance to that
possibility of unification.  Moreover,  its absence would  be
a~very strange fact.  Let  us  add some  more details.

The manner of extending the~flat  space  Dirac  equation  to
general relativity case indicates clearly that the~Lorentz  group
underlies equally both these theories. In other words, the~Lorentz
group retains its importance and significance  at  changing  the
Minkowski space  model  to  an~arbitrary  curved  space-time.  In
contrast to this,  at  generalizing  the~Proca   formulation,  we
automatically destroy any relations to the~Lorentz group, although
the~definition itself for a~spin $1$ particle as an~elementary
object was  based on    this  group.  Such  a~gravity
sensitiveness to the~fermion-boson division might appear rather
strange  and  unattractive asymmetry, being subjected to
the~criticism. Moreover,  just  this feature  has brought about
a~plenty  of  speculations  about  this matter. In any case, this
peculiarity of particle-gravity  field interaction  is  recorded
almost  in  every  handbook on  fiels in curved space-times.

\section{Separation of variables}

In contrast to most of approaches used in the literature and  based on group theoretical arguments
we start our analysis of the  spin 1  field  with the use of the old and conventional  tetrad
formalism   of Tetrode-Weyl-Fock-Ivanenko  applied to matrix Duffin -- Kemmer -- Petiau formalism (see \cite{Book-2009}).
With the use of diagonal static spherical tetrad in anti de Sitter
space-time  $x^{\alpha} = (t,r,\theta, \phi)$  \cite{Hawking-Ellis-1973}
and corresponding Ricci coefficients
\begin{eqnarray}
 dS^{2} = \;(1 + r^{2}) \;  dt^{2} - {dr^{2} \over 1 + r^{2}} -
r^{2} (d\theta^{2} +  \sin ^{2}\theta d\phi ^{2}) \; \; ,
\nonumber
\\
\Phi = 1 +r^{2} \; , \;\;  g_{\alpha\beta}=\left |
\begin{array}{cccc}
                  \Phi  &  0       &  0       &   0 \\
                  0  &  -1/\Phi      &  0       &   0 \\
                  0  &  0       &  - r^{2}  &  0  \\
                  0  &  0       &  0   &   - r^{2} \sin^{2}{\theta}
                        \end {array}
                \right | ,
\nonumber
\\
e^{\alpha}_{(0)}=({1 \over  \sqrt{\Phi} }, 0, 0, 0) \; , \qquad
e^{\alpha}_{(3)}=(0, \sqrt{\Phi}, 0, 0) \; , \nonumber
\\
e^{\alpha}_{(1)}=(0, 0, \frac {1}{r}, 0) \; , \qquad
e^{\alpha}_{(2)}=(1, 0, 0, \frac{1}{ r  \sin \theta})  \; ,
\nonumber
\\
\gamma_{030} ={ r \over \sqrt{ \Phi }} \; , \; \gamma_{311} ={
\sqrt{ \Phi } \over r} \; , \; \gamma_{322} ={ \sqrt{ \Phi }
\over r} \; , \; \gamma_{122} ={ \cos \theta \over r \sin
\theta} \; ,
\end{eqnarray}

\noindent we get to explicit form of a matrix  Duffin -- Kemmer
equation  for a massive spin 1 particle
\begin{eqnarray}
 [ \;i \beta ^{0} \partial _{t}  + i
\Phi ( \beta ^{3}
\partial _{r}  +  {1 \over r}  ( \beta ^{1} j^{31}  +
\beta ^{2} j^{32} ) + {\Phi' \over 2 \Phi } \beta ^{0} J^{03} )
\nonumber
\\
+
{ \sqrt{\Phi } \over r}\;\Sigma _{\theta,\phi } - m
\sqrt{\Phi }  \; ] \; \Phi (x)  =  0 \; , \nonumber
\\
 \Sigma_{\theta ,\phi } =  \; i\; \beta ^{1}
\partial _{\theta } \;+\; \beta ^{2}\; {i\partial \;+\;i\;j^{12}\cos
\theta  \over \sin \theta  } \; .
\label{1.1}
\end{eqnarray}

\noindent Spherical  waves with $(j,m)$ quantum numbers
should be constructed within the following general
 substitution (we adhere notation developed in
 Red'kov \cite{Red'kov-1998(3), Red'kov-1998(4), Red'kov-1999(1),
 Red'kov-1999(2)}; before similar  techniques  was applied
by Dray \cite{Dray-1985, Dray-1986},
Krolikowski and Turski
\cite{Krolikowski-Turski-1986},
Turski \cite{Turski-1986}
;  many years ago such a tetrad basis was  used by
 Schr\"{o}dinger \cite{Schrodinger-1938}
 and Pauli \cite{Pauli-1939} when looking at the problem
of single-valuedness of wave  functions in quantum theory -- then the case of spin  $S=1/2$ particle was
 specified; transition to spin 1 case is achieved  in (\ref{1.1}) trough  a formal change of basic Dirac
  matrices into Duffin --Kemmer  ones)
 \begin{eqnarray}
\Phi _{\epsilon jm}(x)  = e^{-i\epsilon t}  \; [\; f_{1}(r) \;
D_{0} , \; f_{2}(r) \; D_{-1} , \; f_{3}(r) \; D_{0} , \; f_{4}(r)
\; D_{+1} ,
\nonumber
\\
 \; f_{5}(r) \; D_{-1} , f_{6}(r) \; D_{0} , \; f_{7}(r) \; D_{+1} ,
f_{8}(r) \; D_{-1} , \; f_{9}(r)\;D_{0} ,\; f_{10}(r)
\;D_{+1}\;]\; ;
 \label{1.2}
\end{eqnarray}

\noindent
symbol $D_{\sigma}$ designates Wigner  \cite{Wigner-1927} functions
$D_{-m,\sigma}^{j}(\phi, \theta, 0)$ (we use notation according to the book \cite{Varshalovich-Moskalev-Hersonskiy-1975}, ).
In the literature equivalent techniques of  spin-weighted harmonics
Goldberg et al  \cite{Goldberg-Macfarlane-Newman-Rohrlich-Sudarshan-1967} (se also in
\cite{Penrose-Rindler-1984})
 preferably
is used though equivalence of both approach is known \cite{Dray-1985, Dray-1986}.
 Requirement to diagonalize parity operator,  $ \hat{P}\; \Phi _{\epsilon jm} =
P\; \Phi  _{\epsilon jm} $, gives
\begin{eqnarray}
P = (-1)^{j+1} \; , \qquad f_{1} = f_{3} = f_{6} = 0 \; , \; f_{4}
= - f_{2}\;,
\nonumber
\\
 f_{7} = - f_{5}\;,\; f_{10} = + f_{8}\; ;
\nonumber
\\
P = (-1)^{j} \; , \qquad f_{9} = 0\; , \; f_{4} = + f_{2}\;, \;
f_{7} = + f_{5}\; , \; f_{10} = - f_{8}\; .
\label{1.4}
\end{eqnarray}

\noindent  After separation of variables
(for recursive relations needed see Section {\bf 10}) we arrive to the radial systems
 $(\nu  = \sqrt {j (j + 1)} /2 \; )$:

\vspace{2mm} $ P = (-1)^{j+1} \; ,  \qquad $
\begin{eqnarray}
i \epsilon \; f_{5} \;+\; i \Phi({d \over dr} + {1 \over r}\;+\;
{\Phi '\over 2\Phi})\; f_{8} \;+\; i\nu  {\sqrt{\Phi} \over r} \;
f_{9} \;-\; m\; \sqrt{\Phi }\;  f_{2}  = 0 \; ,
\nonumber
\\
 i \epsilon
\; f_{2} \;-\;m\sqrt{\Phi}\; f_{5} = 0 \; , \qquad
 - i \Phi \;({d \over dr} + {1 \over r})\; f_{2}\;-\;
 m \sqrt{\Phi}\; f_{8} = 0 \; ,
 \nonumber
\\
i 2\nu  {\sqrt{\Phi}\over r}\; f_{2}\;-\; m \;\sqrt{\Phi
}\; f_{9} = 0\; ;
\label{1.5a}
\end{eqnarray}

$ P = (-1)^{j} \; , $
\begin{eqnarray}
 \Phi \; ( {d \over dr} +
{2 \over r}) \;F_{6} \;+\;{2\nu \over r}\; F_{5}\; +\; m \; F_{1}
= 0   \; ,
\nonumber
\\ i \epsilon  \; F_{5} \; +\;  i \Phi\;  ( {d \over
dr} + {1 \over r}) \; F_{8} \;- \; m \; \Phi \; F_{2}  = 0  \; ,
\nonumber
\\
i \epsilon  F_{6} -  i {2 \nu r}  F_{8} - m  F_{3} =
0 \; , \;
 - i \epsilon  F_{2} +  {\nu \over r}
F_{1}  -  m  F_{5} = 0 \; ,
\nonumber
\\
i \epsilon   F_{3}  +  \Phi {d \over dr} F_{1} + m
\;\Phi\; F_{6} = 0\; ,
\nonumber
\\
 i \Phi \; ( {d \over dr} + {1 \over r})
\; F_{2} \; + \; i {\nu \over r} \; F_{3}\;+\; m \; F_{8} = 0 \; ;
\label{1.5b}
\end{eqnarray}

\noindent in (\ref{1.5b})  we have used substitutions
\begin{eqnarray}
 F_{1} =
\sqrt{\Phi} \; f_{1} \; ,
 \; F_{2} = f_{2}\; , \; F_{3} =
\sqrt{\Phi }\; f_{3}\; ,
\nonumber
\\
 F_{5} = \sqrt{\Phi }\; f_{5}\; , \;
F_{6} = f_{6} \; ,\; F_{8} = \sqrt{\Phi}\; f_{8} \; .
\nonumber
\end{eqnarray}

The case  of minimal value
$j=0$ is to be treated separately,  because one must use a special substitution from the very beginning
\begin{eqnarray}
\Phi _{\epsilon jm}(x)  = e^{-i\epsilon t}  \; (\; f_{1}   , \;  0
, \; f_{3}   , \; 0 ,
 \; 0  , f_{6}  , \; 0 , \; 0  , \; f_{9}   ,\; 0 \; )\; ;
\label{1.6a}
\end{eqnarray}

\noindent The angular part of the wave operator  $\Sigma_{\theta,\phi}$  acts as a zero operator
and eq.  (\ref{1.1})  takes the  form
\begin{eqnarray}
[ \; i \beta ^{0} \partial _{t}  + i \Phi ( \beta ^{3}
\partial _{r}  +  {1 \over r}\; ( \beta ^{1} j^{31} +
\beta ^{2} j^{32} )
\nonumber
\\
+
 {\Phi' \over 2 \Phi }  \beta ^{0} J^{03} )  - m
\sqrt{\Phi }\;   ] \; \Phi (x)  =  0 \; ;
\label{1.6b}
\end{eqnarray}

\noindent  correspondingly we have a very simple radial system
\begin{eqnarray}
-\Phi \; ({d \over dr}  + {2 \over r} ) \; f_{6}   - m \sqrt{\Phi
} \; f_{1}  = 0 \; , \qquad i\epsilon f_{6}  - m \sqrt{\Phi }
\;f_{3} = 0 \; ,
\nonumber
\\
- i \epsilon  f_{3}  - \Phi  ({d \over dr} + {\Phi '\over 2\Phi
})\; f_{1}  - m \sqrt{\Phi} \;f_{6} = 0 \; , \qquad
 f_{9} = 0 \; .
\label{1.6c}
\end{eqnarray}

\noindent System  (\ref{1.6c})  describes states with parity  $P =
(-1)^{0}=+1$; states with   $P = (-1)^{0+1}=-1$ do not exist.
The system   (\ref{1.6c}) reduces to second order differential equation for  $f_{6}$:
\begin{eqnarray}
{d^{2} \over dr^{2}} f_{6}\; +\; {2 (1 + 2r^{2}) \over r (1 +
r^{2}) } {d \over d r} f_{6} \;
\nonumber
\\
+  \left [ {\epsilon ^{2} \over (1
+ r^{2})^{2}}\; -\; { m^{2} - 2 \over 1 + r^{2} } \;-\; { 2 \over
r^{2}(1+ r^{2}) } \right ] \;  f_{6} = 0 \; ,
\label{1.8}
\end{eqnarray}

\noindent which  is solved in hypergeometric functions
\begin{eqnarray}
f_{6} (r) = r\; (1+ r^{2})^{-\epsilon/2} \;F(\alpha, \beta,
\gamma, -r^{2})\; ,
\nonumber
\\
 \gamma = 1+ 3/2 \; ,
 \qquad \alpha = {3/2 +1   - \epsilon + \sqrt{ m^{2} +1/4} \over 2}\; ,
\nonumber
\\
\qquad \beta = { 3/2 +1 - \epsilon - \sqrt{ m^{2} +1/4} \over 2}\;
. \label{1.9b}
\end{eqnarray}

\noindent The hypergeometric series becomes a polynomial when
 $ \alpha = -n\;, \; n = 0, 1, 2 , ...$; so we arrive at an energy quantization rule
 \begin{eqnarray}
\epsilon =
  N +  3 / 2 + \sqrt{m^{2} + 1 / 4}  \; , \qquad N = 2n + 1 \in \{ 0,1,2, \; \} \; .
\label{1.9c}
\end{eqnarray}

\section{Solutions of radial equations at  $j>0$}

Let us turn to eqs. (\ref{1.5a}). Expressing  $f_{5}, f_{8}, f_{9}$
through  $f_{2}$
\begin{eqnarray}
f_{5} = {i\over m \sqrt{\Phi}} \; \epsilon \;f_{2} \;\;,\qquad
f_{9} = {i \over m}\; {2\nu \over
r}\; f_{2}\; ,
\nonumber
\\
f_{8} =- {i\over m \sqrt{\Phi}}\; \Phi\;({d \over dr}\;+\; {1
\over r})\; f_{2} \;,
\label{2.1a}
\end{eqnarray}

\noindent for  $f_{2}$ we get
\begin{eqnarray}
{d^{2} \over dr^{2}} f_{2}  +  {2 (1 + 2r^{2}) \over r (1 +
r^{2}) } {d \over d r} f_{2} +
 \left [ {\epsilon ^{2} \over (1
+ r^{2})^{2}} -  { m^{2} - 2 \over 1 + r^{2} }- { j(j + 1)
\over r^{2}(1+ r^{2}) } \right ]   f_{2} = 0 \; .
\nonumber
\\
\label{2.1b}
\end{eqnarray}

\noindent Below solutions of this type are referred as $j$-waves
Eq.  (\ref{2.1b})  is solved in hypergeometric functions
\begin{eqnarray}
f_{2}=U_{\epsilon,j}= (-z)^{j/2}   (1-z)^{-\epsilon /2}
F(\alpha  , \beta , \gamma ;   z )\;, \qquad \gamma = j +3/2 \;
,
\nonumber
\\
\alpha = { 3/2 + j - \epsilon + \sqrt{ m^{2} +1/4} \over 2}\; ,
\qquad \beta = { 3/2 + j - \epsilon - \sqrt{ m^{2} +1/4} \over
2}\;   .
\nonumber
\\
\label{2.1d}
\end{eqnarray}

\noindent Restriction  $ \alpha = -n\;, \; n = 0, 1, 2 , ...$ makes hypergeometric series
  polynomials, so we get a quantization rule for energy levels:
\begin{eqnarray}
\epsilon =   N +  {3/2} + \sqrt{m^{2} +1/4}  \; , \qquad N = 2n +
j \in \{ 0,1,2, \; \} \; .
\label{2.1e}
\end{eqnarray}

\noindent It is verified easily that ar $z \rightarrow -\infty $ the radial  function  $U_{\epsilon,j}(z)$
tends ro zero
\begin{eqnarray}
U_{\epsilon,j} (z \rightarrow  -\infty ) \sim z^{j /2} \;
z^{-\epsilon / 2} z^{n}
 \sim
  z^{-3/4 - \sqrt{m^{2} +1/4}\; /2 } \; .
\nonumber
\end{eqnarray}

Now let us turn to eqs.  (\ref{1.5b}). With the use of two different substitutions
\begin{eqnarray}
I. \qquad F_{1} =   \sqrt{j + 1} \; G_{1} \;,\; F_{2} = i \sqrt{
j/2} \; G_{2} \;, \; F_{3} = i \sqrt{J + 1} \; G_{3} \; ,
\nonumber
\\
\qquad F_{5} = \sqrt{j/2} \; G_{5}\; ,\; F_{6} = \sqrt{j + 1} \;
G_{6} \; , \; F_{8} = \sqrt{j/2} \; G_{8}\; ;
\label{2.2a}
\end{eqnarray}
\begin{eqnarray}
II. \qquad F_{1} = \sqrt{j} \; G_{1}\; , \; F_{2} = i
\sqrt{(j+1)/2}\; G_{2}\; , \; F_{3} = i\sqrt{j} \; G_{3}\; ,
\nonumber
\\
\;\; F_{5}  = \sqrt{(j + 1)/2}\; G_{5}\; , \; F_{6} = \sqrt{j}
\;G_{6} \; ,\; F_{8} = \sqrt{(j+1)/2} \; G_{8}\; ,
\label{2.2b}
\end{eqnarray}

\noindent and expressing  $G_{5}, \;G_{6},\; G_{8}$ through $G_{1},\; G_{2},\; G_{3}$ we arrive at
three equations respectively

\vspace{2mm} $ I. \;\; $
\begin{eqnarray}
(  {j(j+1) \over r^{2} } \;+\; m^{2}\; -\; \Phi\;
 ({d \over dr} + {2 \over r})\;
 {d \over dr} ) \;G_{1}
 \nonumber
 \\
 +  \;
{ \epsilon j \over r } \; G_{2} \;+\; \epsilon \; \Phi \; ({d
\over dr} \;+\; {2 \over r})\;
 {1 \over \Phi } \; G_{3} = 0 \; ,
\nonumber
\\
(\epsilon ^{2} \;- \;m^{2} \; \Phi ^{2} \;+\; \Phi ({d \over dr} +
{1 \over r}) \;\Phi\; ({d \over dr}
\nonumber
\\
+ \; {1 \over r} )  )\; G_{2}
\;+\;
 {\epsilon \; (j+1) \over r} \; G_{1} \;+\; \Phi\; {j+1 \over r} \;
 {d  \over dr} \; G_{3} = 0 \;,
\nonumber
\\
({\epsilon ^{2} \; \over \Phi } \;-\; ( {j(j+1) \over r^{2} }
\;m^{2} )\; G_{3} \;-
\nonumber
\\
 - \;\epsilon \;{d \over dr}\; G_{1} \;- \; {j
\over r} \; \Phi \; ({d \over dr} \;+\; {1 \over r} ) \; G_{2} = 0
\; ;
\label{2.4a}
\end{eqnarray}

$ II.  \;\; $
\begin{eqnarray}
( {j(j+1) \over r^{2} } \;+\; m^{2} \;-\; \Phi \; ({d \over dr}
\;+\;{2\over r} )  \; {d \over dr}) G_{1} \;+\;
\nonumber
\\
+ {\epsilon (j+1)
\over r}\; G_{2} \;+\; \epsilon\; \Phi\; ({d \over dr} + {2 \over
r} )
 \; {1 \over \Phi}\; G_{3} = 0 \; ,
\nonumber
\\
( \epsilon ^{2} \;-\; m^{2} \Phi ^{2} \;+\; \Phi\; ({d \over dr}
 \;+\;{1 \over r} )\; \Phi \; ({d \over dr} \;+\;
  {1 \over r})  )\; G_{2}
  \nonumber
  \\
  + \;
{ \epsilon  j \over r}\; G_{1} \;+\; \Phi \; {j \over r} \; {d
\over dr} \; G_{3} = 0 \; ,
\nonumber
\\
( {\epsilon ^{2} \over \Phi } \;- \; {j(j+1) \over r^{2} } \;-\;
m^{2}  )\; G_{3}
\nonumber
\\
- \;
 \epsilon\; {d \over dr} \;G_{1} \;-\; {(j+1) \over r}
\; \Phi \; ({d \over dr} \;+\; {1 \over r} )\; G_{2} = 0 \; .
\label{2.4b}
\end{eqnarray}

To solve eqs.  (\ref{2.4a})  and (\ref{2.4b}), one can make use of the Lorentz condition.
Its explicit form can easily found
\begin{eqnarray}
{- i \epsilon \over \sqrt{\Phi }}\; f_{1} -  \sqrt{\Phi }\; ({d
\over dr}  +
 {2 \over r} + {\Phi ' \over 2 \Phi } )\; f_{3} -  { \nu  \over r}\; ( f_{2} + f_{4} ) = 0 \;.
\label{2.5a}
\end{eqnarray}

\noindent When  $P = (-1)^{j+1}$ eq. (\ref{2.5a})  holds identically; for substitutions I    and    II
in  (\ref{2.2a})  and   (\ref{2.2b}) it gives respectively
\begin{eqnarray}
I. \qquad   - \epsilon \; {G_{1} \over \Phi } = {j \over r}\;
G_{2} \;+\; ({d \over d r } \;+ \;{2 \over r}) \; G_{3} \; ,
\nonumber
\\
II. \qquad -\epsilon\;{G_{1} \over \Phi } = {j +1 \over r}\; G_{2}
\;+ \; ({d \over d r } \;+\; {2 \over r}) \; G_{3} \; .
\label{2.5c}
\end{eqnarray}

\noindent Allowing for relations  (\ref{2.5c}), let us express $G_{1}$  through
$G_{2}$  and  $G_{3}$,  and substitute the results into 2-nd and 3-d
equations in (\ref{2.4a}) and  (\ref{2.4b}). Thus we get
respectively

$ I. $
\begin{eqnarray}
[ {d^{2} \over dr^{2} } \;+\; ({2 \over r} \;+\; {\Phi ' \over
\Phi} ) {d \over d r } \;+\; {\Phi' \over r \Phi } \;+\;{\epsilon
^{2} \over \Phi ^{2}}
\nonumber
\\
- {m^{2} \over \Phi } \;-\; {j(j+1)
\over \Phi r^{2} }  ]\; G_{2} \;-\; {2 (j + 1) \over r^{2} \Phi}
\;  G_{3} = 0  \; ,
\nonumber
\\
\; [ {d^{2} \over dr^{2} } \;+\; ({2 \over r} \; +\; {\Phi ' \over
\Phi } )\; {d \over d r } \;+\; {2 \Phi ' \over r \Phi }\;-\;{ 2
\over r^{2} }\; +\; {\epsilon ^{2} \over \Phi ^{2}}
\nonumber
\\
- \; {m^{2}
\over \Phi } \;-\; {j(j+1) \over \Phi  r^{2}} ] \; G_{3} \;-\; {2
j \over r^{2} \Phi}\; G_{2} = 0 \; ;
\label{2.6a}
\end{eqnarray}

$ II. $
\begin{eqnarray}
\; [ {d^{2} \over dr^{2} } \;+\; ({2 \over r} \;+\; {\Phi ' \over
\Phi }) {d \over d r }\; + \; {\Phi ' \over r \Phi } \;+ \;
{\epsilon ^{2} \over \Phi ^{2}}
\nonumber
\\
- \;
 {m^{2} \over \Phi } \;- \;{j(j+1) \over  \Phi r^{2} } ]
  \; G_{2}
\; - \; {2 j \over r^{2} \; \Phi }\; G_{3} = 0 \;,
\nonumber
\\
\; [ {d^{2} \over dr^{2}} \;+\; ({2 \over r} \;+\; {\Phi '\over \Phi
}) \; {d\over d r} \;+\; {2 \Phi ' \over r \Phi}\; -\;{ 2 \over
r^{2}} \; + \; {\epsilon ^{2} \over \Phi ^{2}}
\nonumber
\\ -
 {m^{2}\over \Phi }\; - \;
{j(j+1) \over \Phi r^{2}} ]\; G_{3} \;-\; {2 (j+1) \over r^{2}
\Phi}\; G_{2} = 0 \; .
\label{2.6b}
\end{eqnarray}

In the case  $I$,  taking  $G_{3} = + G_{2}$, from two equations (\ref{2.6a}) we get one the same
\begin{eqnarray}
I. \qquad   G_{3} = + G_{2}=U_{\epsilon, j+1}  \; , \qquad [ \;
{d^{2} \over dr^{2} } \;+\; {2 (1 + 2r^{2}) \over
 r (1 + r^{2})}\; {d \over dr }
\nonumber
\\
+ \;  { \epsilon ^{2} \over (1 + r^{2})^{2} } \; - \; {M^{2} \;-\;
2 \over 1 + r^{2}} \;-\; {(j+1)(j+2)  \over r^{2}(1 + r^{2}) } \;
]\; G_{2}  = 0 \;  .
\label{2.7a}
\end{eqnarray}

\noindent In the same manner, in the case   II, taking $G_{3} =  -
G_{2}$,  we get one the same equation (it differs from previous one
by the simple formal changing  $(j + 1)$  into  $(j - 1)$:
\begin{eqnarray}
II. \qquad   G_{3} = - G_{2}=U_{\epsilon, j-1} \qquad [ \; {d^{2}
\over dr^{2} } \;+\; {2 (1 + 2r^{2}) \over
 r (1 + r^{2})}\; {d \over dr }
\nonumber
\\
+ \;  { \epsilon ^{2} \over (1 + r^{2})^{2} } \; - \; {M^{2} \;-\;
2 \over 1 + r^{2}} \;-\; {(j-1)j)  \over r^{2}(1 + r^{2}) } \; ]\;
G_{2}  = 0 \;  .
\label{2.7b}
\end{eqnarray}

Thus, beside the waves of $j$-type, there exist else two types
(all technical details of calculations with hypergeometric function are omitted)

\vspace{2mm} $ I. \qquad   (j + 1)  - \mbox{type}\; , $
\begin{eqnarray}
 G_{3} =
G_{2} = U_{\epsilon, j+1}  \; , \qquad
   - \epsilon \; {G_{1} \over \Phi } =  ({d \over d r } \;+ \;{j+2 \over r}) \; G_{2}
\; ;
\nonumber
\\
G_{1} =
  \sqrt{-z} \;\;
 U_{\epsilon, j+1}  - { 2j+3 \over \epsilon} \; \sqrt{1-z} \;  \; U_{\epsilon -1, j}  \; ,
\nonumber
\\
U_{\epsilon,j+1} =  (-z)^{(j+1)/2}  \; (1-z)^{-\epsilon /2} \;
F(\alpha  , \beta , \gamma ;  \; z )\;,
\nonumber
\\
U_{\epsilon -1 ,j} = (-z)^{j/2}  \; (1-z)^{-(\epsilon -1) /2} \;
F(\alpha  , \beta  , \gamma -1  ;  \; z )\; ,
\nonumber
\\
\gamma = j+1  +3/2 \; , \qquad \alpha = { 3/2 + j +1 - \epsilon + \sqrt{ m^{2} +1/4} \over 2}\; ,
\nonumber
\\
 \beta = { 3/2 + j+1  - \epsilon - \sqrt{ m^{2} +1/4} \over 2}\;  ;
\label{2.8}
\end{eqnarray}

$ II. \;\;\;  (j - 1) - \mbox{type} $,
\begin{eqnarray}
 -G_{3} =  G_{2} =
U_{ \epsilon , j-1} \; , \qquad -\epsilon\;{G_{1} \over \Phi } =
(- {d \over d r } \;+\; {j-1 \over r}) \; G_{2} \;  ,
\nonumber
\\
G_{1} = -\sqrt{-z} \;\;
 U_{\epsilon, j-1}  - { 2 \over \epsilon} \; { \alpha \beta \over \gamma }\;\sqrt{1-z} \;  \; U_{\epsilon -1, j}  \; ,
\nonumber
\\
U_{\epsilon ,j-1}  = (-z)^{(j-1)/2}  \; (1-z)^{-\epsilon /2} \;
F(\alpha , \beta , \gamma ;  \; z ) \; ,
\nonumber
\\
U_{\epsilon -1 ,j}  = (-z)^{j/2}  \; (1-z)^{-(\epsilon -1) /2} \;
F( \alpha+1   ,  \beta +1   , \gamma +1   ;  \; z )\; ,
\nonumber
\\
\gamma = j-1  +3/2 \; , \qquad
\alpha  = { 3/2 + j -1 - \epsilon + \sqrt{ m^{2} +1/4} \over 2}\;
,
\nonumber
\\
\beta = { 3/2 + j -1 - \epsilon - \sqrt{ m^{2} +1/4}
\over 2}\; .
 \label{2.9}
 \end{eqnarray}

Let us collect results together. There are constructed solutions f three types
(below only  $f_{1},\ldots ,f_{4}
$ are specified):

\vspace{5mm}

\underline{$ j - \mbox{wave}$}
\begin{eqnarray}
  f_{1} = f_{3} = 0 \;\;
, \; \; f_{2} = - f_{4} = U_{ \epsilon ,j} \; ;
\nonumber
\label{13.2.13a}
\end{eqnarray}

\underline{$ (j + 1) - \mbox{wave} $},
\begin{eqnarray}
f_{1} = \sqrt{j+1} \; \;\left [
  {\sqrt{-z} \over \sqrt{1-z} }  \;\;
 U_{\epsilon, j+1}  - { 2j+3 \over \epsilon} \; \; U_{\epsilon -1, j} \; \right  ]  \;  ,
\nonumber
\\
f_{2} = + f_{4} = + i \; \sqrt{j/2} \; \; U_{\epsilon ,j+1} \;\; ,
\qquad \; f_{3} = + i \;\sqrt{j+1}\;  { 1\over \sqrt{1 - z} } \;
\; U_{\epsilon ,j+1}  \; ;
\label{2.13b}
\nonumber
\end{eqnarray}

\underline{$ (j - 1) - \mbox{wave} \; ,$}
\begin{eqnarray}
f_{1} = \sqrt{j}\; \left  [\;   -{\sqrt{-z}  \over \sqrt{1 -z} }
\;\;
 U_{\epsilon, j-1}  -
 { 2 \over \epsilon} \; { \alpha \beta \over \gamma }\; \; U_{\epsilon -1, j}  \; \right  ] \; ,
\nonumber
\\
f_{2} = + f_{4} = i\;\sqrt{{j+1\over 2}} \; U_{\epsilon ,j-1} \; ,
\qquad  f_{3} = - i\;\sqrt{j} {1 \over \sqrt{1 -z}}\;\;
U_{\epsilon ,j-1}\; .
\nonumber
\\
\label{2.13c}
\end{eqnarray}

\noindent Three types of solutions correspond to three possible values
of the  orbital angular moment for spin 1 particle at fixed $j: \; l = j , j + 1 , j - 1$ .

\section{Massless limit for spin 1  particle}

Let us shortly consider a massless limit. The Duffin -- Kemmer  -- Petiau  equation (\ref{1.1})
stays much the same with only formal change
\begin{eqnarray}
 m \; \sqrt{\Phi } \;\; \rightarrow  \;\;
 P_{6} \;  \sqrt{ \Phi} \;, \qquad
P_{6} = \left | \begin{array}{cccc}
0 & 0 & 0 & 0 \\  0 & 0 & 0 & 0 \\
0 & 0 & I & 0 \\  0 & 0 & 0 & I \end{array} \right |
\label{3.1}
\end{eqnarray}

\noindent which produces evident alterations  in the radial system
\begin{eqnarray}
m \sqrt{\phi } f_{i} \;\;\rightarrow \;\; 0 \; , \;\; \mbox{at}\;\;
 i = 1, 2, 3, 4 \; ;
 \nonumber
 \\
 m \sqrt{\phi } f_{i} \;\;\rightarrow  \;\;
\sqrt{\phi} f_{i}\; , \;\;
\mbox{at}  \;\; i = 5, \ldots 10 \; .
\label{3.1b}
\end{eqnarray}

\noindent In massless case the Lorentz  condition must be considered
as an external gauge restriction for photon field.
Other relations   remain the same,  instead of old parameters of hypergeometric
functions now one must  take  new ones
\begin{eqnarray}
U_{\epsilon,j}\;, \qquad \alpha = { 2 + j - \epsilon  \over 2}\; ,
\qquad  \beta = { 1 + j - \epsilon  \over 2}\; , \qquad
\gamma = j +3/2 \; ;
\label{3.2a}
\end{eqnarray}

\noindent the energy quantization rule looks
\begin{eqnarray}
\epsilon = 2n + j + 2 =   N +  2  \; , \qquad N = 2n + j \in \{
0,1,2, \; \} \; . \label{3.2b}
\end{eqnarray}

The case of minimal value $j=0$ should be considered separately.
The system  (\ref{1.6c}) becomes
\begin{eqnarray}
 -\Phi \; ({d \over dr}  +
{2 \over r} ) \; f_{6}   - 0 \; \sqrt{\Phi } \; f_{1}  = 0 \; ,
\qquad i\epsilon f_{6}  - 0\; \sqrt{\Phi } \;f_{3} = 0 \; ,
\nonumber
\\
- i \epsilon  f_{3}  - \Phi  ({d \over dr} + {\Phi '\over 2\Phi
})\; f_{1}  -  \sqrt{\Phi} \;f_{6} = 0 \; , \qquad
 f_{9} = 0 \; ,
\label{3.3}
\end{eqnarray}

\noindent which is equivalent to
\begin{eqnarray}
g_{6}=0\;, f_{9} = 0\; , \qquad - i \epsilon  f_{3}  - \Phi  ({d
\over dr} + {\Phi '\over 2\Phi })\; f_{1}   = 0 \; ,
\label{3.4}
\end{eqnarray}

\noindent Therefore, for all states of electromagnetic field at
$j=0$  the components of electric and magnetic vectors vanish
($F_{\alpha \beta}=0$); and non-vanishing  $f_{1}(r), f_{3}(r)$  correspond to
solutions of gradient type
  $A_{\alpha} = \nabla_{\alpha} \Phi$.
To have fixed two radial functions in (\ref{3.4}), one must impose certain gauge condition.
In particular,  taking the Lorentz condition
   (see  (\ref{2.5a})), we get equations
 (let it be  $
f_{1} = \Phi^{-1/2}  F_{1} , \; f_{3} = \Phi^{-1/2}  F_{3}$):
\begin{eqnarray}
-{i\epsilon \over \Phi}\; F_{3} - {d \over dr} \; F_{1} = 0 \;,
\qquad -{i\epsilon \over \Phi}\; F_{1} - ( {d \over dr} +{2 \over
r} )\; F_{3} = 0 \; ;
\label{3.6}
\end{eqnarray}

\noindent from whence it follows
\begin{eqnarray}
\left[{d^{2}\over dr^{2}}+{2 (1 +2r^{2}) \over r (1+r^{2})}{d\over
dr}+{\epsilon^2 \over (1+r^{2})^{2}}\right]F_{1}=0 \; ,
\label{3.7}
\end{eqnarray}

\noindent which represent  $j=0$-spherical solution of the equation
 $ \nabla^{\alpha} \nabla_{\alpha} \Phi =0 , \; \Phi =
e^{-i\epsilon t} f(r) $.

\section{ 5-dimensional form of the wave equation   }

It is well known  that wave equation for a particle with spin 1 in the de  Sitter and anti de Sitter spaces
can be presented in 5-dimensional form explicitly invariant under  groups
$SO(4.1)$ and $SO(3.2)$ respectively.
Let us specify the problem of spherical wave solutions in anti de Sitter model to
the  5-dimensional formalism.
It is convenient to start with conformal  flat coordinates
\begin{eqnarray}
dS^{2} =  { (dx^{0})^{2}  - (dx^{1})^{2}
- (dx^{2})^{2} - (dx^{3})^{2} \over \Phi^{2} }
\; ,\;\;
\Phi  = (1 + x^{2})/2 \;  ;
\label{1.1'}
\end{eqnarray}

\noindent the Proca  tensor equations read
\begin{eqnarray}
\partial _{\alpha } \Psi _{\beta }  - \partial _{\beta } \Psi _{\alpha } =
m\; \Psi _{\alpha \beta } \;, \qquad  \Phi ^{2} \; \partial
^{\beta } \Psi _{\alpha \beta } = m \Psi _{\alpha } \; .
\label{1.2'}
\end{eqnarray}

\noindent
Let us introduce five coordinates  $\xi ^{a}\; (a = \alpha  ,\;   5
)$:
\begin{eqnarray}
\xi ^{\alpha } = { x^{\alpha } \over \Phi  }\; , \qquad  \xi ^{5} = { 1 - x^{2} \over 1 + x^{2}}
\; ,
\nonumber
\\
x^{\alpha } = { \xi ^{\alpha } \over 1  + \xi ^{5} }\; , \qquad
\Phi  = { 1 \over 1  + \xi ^{5}} \; ,
\nonumber
\\
 (\xi ^{0})^{2} - ( \xi ^{1})^{2} - ( \xi ^{2})^{2} - ( \xi
^{3})^{2} + ( \xi ^{5})^{2}  = + 1 \;,
\nonumber
\\
dS^{2} = \eta _{\alpha \beta } \; d\xi ^{\alpha } d\xi ^{\beta } \;+\;
(d\xi ^{5})^{2}\; .
\label{1.3'}
\end{eqnarray}

\noindent
In other words, the anti de Sitter space-time can be considered as a
sphere in 5-dimensional pseudo-Euclidean space;
therefore the model permits 10-parametric symmetry group $SO(3.2)$.
Instead  $\Psi ^{\alpha }(x)$ (in the following designated as  $a^{\alpha }(x) )$
let us introduce 5-vector  $A^{a}(\xi)$:
\begin{eqnarray}
 A^{\alpha }  = ( \;
{  \delta ^{\alpha }_{\beta }   \over \Phi } \;-\; {x^{\alpha }\; x_{\beta
}  \over \Phi ^{2}} )\;  a^{\beta }   \; , \;\; A^{5} = -{ x_{\beta } \; a^{\beta } \over \Phi ^{2}}\;
  \;  ,
  \nonumber
  \\
  a^{\alpha }(x) =  \Phi \; ( A^{\alpha }\;  - \; x^{\alpha }
A^{5})\;  .
\label{1.4'}
\end{eqnarray}

\noindent These five components  $A^{a}(\xi )$  obey additional restriction
\begin{eqnarray}
 A^{a} \xi _{a}  =  A^{0} \xi^{0} -\vec{A} \; \vec{\xi} + A^{5} \xi^{5} =  0 \;.
 \nonumber
 \end{eqnarray}

Invariant with respect to  $SO(3.2)$  wave equation for vector  $A^{a}(\xi )$  should be constructed
with the help of  the operator
$
L_{ab}   =  \xi _{a}\; (\partial / \partial \xi ^{b})  - \xi
_{b} \; (\partial  / \partial \xi ^{a} )$
and looks as follows
\begin{eqnarray}
(- {1 \over 2} L^{ab}  L_{ab}  +  m^{2}  -  2 )
A_{c} = 0 \; ,
\nonumber
\\
L_{ab} A^{b}  = A_{a} \; , \qquad
A^{a} \xi _{a} = 0 \; .
 \label{1.6'}
\end{eqnarray}

\section{Spherical waves in 5-dimensional form,  \\ separation of variables}

Let us consider  equations (\ref{1.6'}) in static coordinates of the anti de Sitter space
\begin{eqnarray}
\xi ^{1}  = r \sin \theta  \cos \phi  \; , \; \xi ^{2}  = r \sin
\theta  \sin \phi  , \; \xi^{3} = r \cos \theta  \; ,
\nonumber
\\
 \xi ^{0}
= \sin \;t  \sqrt{1  + r^{2} } \;  , \; \xi ^{5}  = \cos \;t
\sqrt{1 + r^{2}}   \; ;
\nonumber
\\
t  = \mbox{arctg}\; { \xi ^{0} \over \xi ^{5}} \; , \; r  = \sqrt{
(\xi ^{1})^{2}  + (\xi ^{2})^{2}  + (\xi ^{3})^{2} } \; , \;
\nonumber
\\
\theta  = \mbox{arctg} \; { \sqrt{(\xi ^{1})^{2} + (\xi ^{2})^{2}}
 \over \xi ^{3}} \; , \;
\phi  = \mbox{arctg} \; { \xi ^{2} \over \xi ^{1}} \; .
\label{2.2'}
\end{eqnarray}

\noindent
For any representation of the group  $SO(3.2)$ on the functions  $\Psi (\xi )$ we have
\begin{eqnarray}
\xi ' = S\; \xi \; , \;\; \Psi '(\xi ')  = U \; \Psi (\xi ) \qquad
\Longrightarrow \qquad  \Psi '(\xi )  = U\;  \Psi (S^{-1} \; \xi )
\; .
\nonumber
\end{eqnarray}

\noindent In particular case  $U \equiv  S$  and
$\Psi  \equiv  A$, for rotation in the plane $0-5$
\begin{eqnarray}
\xi ^{0'}  = \cos    \omega  \;  \xi ^{0}  -
             \sin  \omega  \;  \xi ^{5} \; , \qquad
 \xi  ^{5'}  = \sin  \omega \;  \xi ^{0}  +
             \cos   \omega  \;  \xi ^{5} \; ,
\nonumber
\end{eqnarray}

\noindent we get
\begin{eqnarray}
A' (\xi ) = ( I  \;+\; \delta  \omega \;  J_{50} )\; A(\xi ) \; ,
\qquad J_{50}  =  {\cal L}_{50}  + \sigma _{50}  \; ,
\nonumber
\\
{\cal L}_{50}  =  \xi _{5}\;{\partial  \over \partial \xi ^{0} }
\;-\; \xi ^{0} \; {\partial \over \partial \xi ^{5} }\; , \qquad
\sigma _{50} = \left | \begin{array}{ccccc}
0 & 0 & 0 & 0 & -1 \\
0 & 0 & 0 & 0 & 0 \\
0 & 0 & 0 & 0 & 0 \\
0 & 0 & 0 & 0 & 0 \\
+1 & 0 & 0 & 0 & 0
\end{array} \right |   \; .
\label{2.3}
\end{eqnarray}

\noindent Generators    will be used to construct  5-form for energy operator, and total quantum moment.
  From requirements
\begin{eqnarray}
( + i J_{50} )^{a}_{\;\;b} \; A^{b}  = \epsilon  A^{a} \; ,
\nonumber
\\
 (\vec{J}^{\;2}) ^{a}_{\;\;b} A^{b} =
 j(j+1)\; A^{a} \; , \qquad ( J_{3})^{a}_{\;\;b} \; A^{b}  =
 m \; A^{a}
\label{2.4'}
\end{eqnarray}

\noindent
it follows
 \begin{eqnarray}
\vec{A}= e^{-i\epsilon t}
 \; \left [ \; f (r) \; \vec{Y}^{j+1}_{jm}(\theta,\phi) + g(r ) \; \vec{Y}^{j-1}_{jm}(\theta,\phi)
  + h(r ) \; \vec{Y}^{j}_{jm}(\theta,\phi) \; \right ] \; ,
\nonumber
\\
A^{0} = \left [  \;  e^{-i(\epsilon-1)t} \; F(r)   +  i \;
e^{-i(\epsilon +1)t} \;  G(r)   \; \right ] \; Y_{jm}(\theta,\phi)
\; ,
\nonumber
\\
A^{5} = \left [ \;  i \;e^{-i(\epsilon-1)t}\;  F(r)  +
e^{-i(\epsilon +1)t} )\; G(r) \; \right ]  \; Y_{jm}(\theta,\phi)
\; .
\label{2.5'}
\end{eqnarray}

\noindent
At given $j=1,2,... $ there exist three linearly independent
spherical vectors  $\nu= j+1,j,j-1$, when $j=0$ there exist only one that
\begin{eqnarray}
j =0\;, \qquad     \vec{A}=
e^{-i\epsilon t}
  \; f (r) \; \vec{Y}^{1}_{00}  \;  ,
  \nonumber
  \\
  {1 \over 2} (A^{0} + i A^{5})  =  i G(r) \;  e^{-i(\epsilon +1)t}   \; ,
  \nonumber
  \\
{1 \over 2} \; ( A^{0} - i A^{5} ) =   F(r) \;
e^{-i(\epsilon -1)t}   \; .
\label{2.6'}
\end{eqnarray}

\noindent
Radial functions $f(r),\; g(r), \;h(r), \; F(r), G(r)$ should be  determined from (\ref{1.6'}).
From the first equation in   (\ref{1.6'}), taking into account action
of  $\vec{l} ^{\; 2}$ on spherical vectors
\cite{Varshalovich-Moskalev-Hersonskiy-1975}
\begin{eqnarray}
\vec{l}^{\; 2} \;  \vec{Y}^{\nu }_{jm} = \nu (\nu +1)\;
\vec{Y}^{\nu }_{jm}\;\;  ,
\nonumber
\\
 \vec{l}^{\; 2} \;  Y_{jm} = j
(j +1)\; Y_{jm} \; , \qquad \nu = j, j+1, j-1 \; ,
\nonumber
\end{eqnarray}

\noindent for radial functions  $f(r), \;  g(r), \;  h(r), \;
F(r)$  we get  equations of one the same type
\begin{eqnarray}
\left [ {d^{2}\over dr^{2}} + {2 (1+2r^{2}) \over r(1+r^{2}) }
{d\over dr} + {\Lambda^{2} \over (1+r^{2}) ^{2} } - {\nu (\nu+1)
\over r^{2} (1+r^{2}) }- {(m^{2}-2)\over 1+r^{2}} \right ]
U_{\Lambda, \nu}=0 \; ;
\nonumber
\\
\label{2.7'}
\end{eqnarray}

\noindent so we know  these five  functions to  within numerical constants
$f_{0}, g_{0},  h_{0}, F_{0},  G_{0}$
\begin{eqnarray}
 f =f_{0} \; U_{\epsilon, j+1} \;  ,  \;\;  g =g_{0} \;  U_{\epsilon, j-1} \;  ,
\;\;  h = h_{0} \; U_{\epsilon, j} \; ,
\nonumber
\\
F = F_{0} \;  U_{\epsilon -1, j}\; , \;\;  G = G_{0} \;
U_{\epsilon +1, j}\; ;
\nonumber
\end{eqnarray}

\noindent these constants should be  obtained from  remaining equations in (\ref{1.6'}).
Solutions of eq.  (\ref{2.7'})  are constructed in terms of hypergeometric functions
 (it suffices to consider in detail only the case
  $U_{\epsilon ,j}$):
\begin{eqnarray}
U_{\epsilon ,j}  = (-z)^{j/2}  \; (1-z)^{-\epsilon /2} \;
F(\alpha  , \beta , \gamma ;  \; z ) \; ,\qquad z=-r^{2}, \;\;  \gamma = j +3/2 \; ,
\nonumber
\\
\alpha = { 3/2 + j - \epsilon + \sqrt{ m^{2} +1/4} \over 2}\; ,
\;\;
 \beta = { 3/2 + j - \epsilon - \sqrt{ m^{2} +1/4} \over
2}\;  ;
\nonumber
\\
\label{2.8'}
\end{eqnarray}

\noindent we have polynomials when
$
\alpha = -n\;, \; n = 0, 1, 2 , ...\;$; which results in the quantization rule for energy levels
\begin{eqnarray}
\epsilon =  N +  {3/2} + \sqrt{m^{2} +1/4}  \; , \qquad N = 2n + j \in \{
0,1,2, \; \} \; .
\label{2.9'}
\end{eqnarray}

\noindent  From two remaining equations in  (\ref{1.6'}) one can produce relationships between
$(G \pm i F)$ and  $(f,g)$:
\begin{eqnarray}
 G- iF   = {1 \over  \epsilon }\;  \sqrt{1 + r^{2} } \;   \left [ \; ( {d \over dr} + {j + 2\over r} )\; f -
 ({d \over dr} - { j - 1 \over r}  )\; g \; \right ]  \;,
 \nonumber
 \\
      G + i F   =  {  -   r f  +     r g  \over  \sqrt{ 1 + r^{2} } }\;    .
\label{2.10'}
\end{eqnarray}

\section{ Solutions of the types
$(j,j+1,j-1)$ }

Let us search three linearly independent solutions in the  form
\begin{eqnarray}
j -\mbox{type} \; , \qquad
 f = 0 \; , \; g = 0 \; ,\;
h \neq  0 \;,
\nonumber
\\
(j+1)-\mbox{type}\; , \qquad
 f \neq  0  \; , \; g = 0 \;
,\; h = 0 \; ,
\nonumber
\\
(j-1)-\mbox{type}\;, \qquad
 f = 0 \; , \; g \neq  0 \;
,\; h = 0 \; .
\nonumber
\end{eqnarray}

\noindent In fact, these requirements are equivalent to diagonalizing of the orbital  angular operator
 $ \vec{l}^{\; \;2} = \nu (\nu  +1)\;, \; \nu
= j+1, j, j -1 $.
First, let us consider the wave  $(j+1)$
\begin{eqnarray}
f= \sqrt{{2j+1 \over j+1}}\; f_{0} \;  U_{\epsilon, j+1} \, ,
\qquad F = F_{0} \; U_{\epsilon -1, j}\; , \qquad  G = G_{0} \;
U_{\epsilon +1, j}\; ; \label{3.1'}
\end{eqnarray}

\noindent eqs.   (\ref{2.10'})
take the form
\begin{eqnarray}
     G + i F   =  - {     r f (r)   \over  \sqrt{ 1 + r^{2} } }\;   , \qquad
 G- iF   = {1 \over  \epsilon }\;  \sqrt{1 + r^{2} } \;  ( {d \over dr} + {j + 2\over r} )\; f  \; .
\label{3.2'}
\end{eqnarray}

\noindent or after transition to the variable
 $z=-r^{2}$
\begin{eqnarray}
2{G_{0} \over f_{0}}  \;  U_{\epsilon +1, j}   = -
 {     \sqrt{-z}        \over  \sqrt{ 1 -z } } \; U_{\epsilon, j+1}  +
{1  \over  \epsilon }\;  \sqrt{1 -z } \;  ( -2 \sqrt{-z} {d \over
d z} + {j + 2\over \sqrt{-z}} )\; U_{\epsilon, j+1}\; ,
\nonumber
\\
2i{F_{0} \over f_{0}}  \;  U_{\epsilon -1, j}   = -
 {     \sqrt{-z}        \over  \sqrt{ 1 -z } } \; U_{\epsilon, j+1}  -
{1  \over  \epsilon }\;  \sqrt{1 -z } \;  ( -2 \sqrt{-z} {d \over
d z} + {j + 2\over \sqrt{-z}} )\; U_{\epsilon, j+1} \; .
\nonumber
\\
\label{3.3'}
\end{eqnarray}

Allowing for  explicit forms
\begin{eqnarray}
U_{\epsilon ,j+1}  = (-z)^{(j+1)/2}  \; (1-z)^{-\epsilon /2} \;
F(\alpha  , \beta , \gamma ;  \; z ) \; ,
\nonumber
\\
U_{\epsilon +1 ,j}
= (-z)^{j/2}  \; (1-z)^{-(\epsilon +1)/2} \;  F(\alpha -1  , \beta
-1 , \gamma -1 ;  \; z ) \; ,
\nonumber
\\
U_{\epsilon -1 ,j}  =  (-z)^{j/2}  \; (1-z)^{-(\epsilon -1) /2} \;  F(\alpha  , \beta
, \gamma -1  ;  \; z )\; ,
\nonumber
\\
\gamma = j+1  +3/2 \; ,
\;\; \alpha = { 3/2 + j +1 - \epsilon + \sqrt{ m^{2} +1/4} \over 2}\; ,
\nonumber
\\
 \beta = { 3/2 + j +1 - \epsilon - \sqrt{ m^{2} +1/4} \over 2}\; ,
  \label{3.4'}
 \end{eqnarray}

\noindent and using known  formulas for
hypergeometric functions we get expressions for  $G_{0},  F_{0}$:
\begin{eqnarray}
G_{0} = {\gamma -1 \over \epsilon } \; f_{0} = {j +3/2 \over
\epsilon }\; f_{0} \; , \qquad
  F_{0}  = i \; {j+3/2  \over  \epsilon } \; f_{0} = i  \; {1 -\gamma  \over  \alpha + \beta - \gamma } \; f_{0}  \; .
 \label{3.5'}
 \end{eqnarray}

Analogous calculations may be performed for the case of $(j-1)$-waves:
\begin{eqnarray}
g=  \sqrt{{2j+1 \over j}}\; g_{0} \; U_{\epsilon, j-1}(z) \; ,
\;\;  F = F_{0} \; U_{\epsilon -1, j}\; , \;\;   G = G_{0} \;
U_{\epsilon +1, j}\; , \nonumber
\\
 G + i F   =  {       r  g   \over  \sqrt{ 1 + r^{2} } }\;   , \;\;
 G - iF   = -  {1 \over  \epsilon }  \sqrt{1 + r^{2} }
 ({d \over dr} - { j - 1 \over r}  )\; g \;  ;
\label{3.6'}
\end{eqnarray}

\noindent or in variable  $z=-r^{2}$
\begin{eqnarray}
2{G_{0} \over g_{0}} \; U_{\epsilon +1, j} = {       \sqrt{-z}
\over  \sqrt{ 1 -z } }\; U_{\epsilon, j-1} -  {1 \over  \epsilon
}\;  \sqrt{1 - z  } \;
 ( -2 \sqrt{-z} {d \over dz} - { j - 1 \over \sqrt{-z}}  )\; U_{\epsilon, j-1} \; ,
\nonumber
\\
2i{F_{0}  \over g_{0}} \; U_{\epsilon -1, j} = {       \sqrt{-z}
\over  \sqrt{ 1 -z } }\; U_{\epsilon, j-1} +  {1 \over  \epsilon
}\;  \sqrt{1 -z } \;
 (-2 \sqrt{-z} {d \over d z} - { j - 1 \over \sqrt{-z}}  )\; U_{\epsilon, j-1} \; .
 \nonumber
 \\
 \label{3.7'}
 \end{eqnarray}

\noindent  Using  the formulas
\begin{eqnarray}
U_{\epsilon ,j-1}  = (-z)^{(j-1)/2}  \; (1-z)^{-\epsilon /2} \;
F(a , b , c ;  \; z ) \; ,
\nonumber
\\
U_{\epsilon +1 ,j}  = (-z)^{j/2}  \; (1-z)^{-(\epsilon +1)/2} \;  F(a  , b , c +1  ;
\; z ) \; ,
\nonumber
\\
U_{\epsilon -1 ,j}  =  (-z)^{j/2}  \; (1-z)^{-(\epsilon -1) /2} \;  F( a+1   ,  b +1
, c +1   ;  \; z )\; ,
\nonumber
\\
 c = j-1  +3/2 \;, \;\; a  = { 3/2 + j -1 - \epsilon + \sqrt{ m^{2} +1/4} \over 2}\; ,
\nonumber
\\
b = { 3/2 + j -1 - \epsilon - \sqrt{ m^{2} +1/4} \over 2}\;
\nonumber
\label{3.8}
\end{eqnarray}

\noindent we arrive at
\begin{eqnarray}
G_{0} = {(a-c)(b-c) \over \epsilon \;  c }\; g_{0}\; , \qquad
 F_{0}  \;
=
  i \;  {a  b\over \epsilon \; c} \; g_{0} \; .
     \;
\label{3.9}
\end{eqnarray}

Collecting together results we have

\vspace{3mm} {\bf $j$-wave} \qquad \underline{$j=1,2,3,...$}
\begin{eqnarray}
  \vec{A} = e^{-i\epsilon t} h_{0} \;
U_{-i\epsilon ,j} (r) \; \vec{Y}^{j}_{jm}(\theta ,\phi ) \; ,
\qquad A^{0} = 0  \; , \; A^{5} =  0  \;     ;
\nonumber
\\
\end{eqnarray}

\noindent
quantization rule $
 \epsilon = 2n + j +3 /2  + \sqrt{m^{2} + 1/  4} $.

\vspace{2mm}

{\bf $(j-1)$-wave }  \qquad \underline{$j=1,2,3,...$}
\begin{eqnarray}
\vec{A}= e^{-i\epsilon t}
 \; \sqrt{{2j + 1 \over j  }} \;  f (r) \; \vec{J}^{j-1}_{jm}(\theta,\phi) \; ,
 \qquad f(r) = f_{0} \; U_{\epsilon, j-1} \; ,
\nonumber
\\
{1 \over 2} (A^{0} + i A^{5})  =  i\;  G(r) \;  e^{-i(\epsilon
+1)t}  \; Y_{jm} \; ,\;\;
{1 \over 2} \; ( A^{0} - i A^{5} ) =   F(r) \;  e^{-i(\epsilon
-1)t}  \; Y_{jm} \;,
\nonumber
\\
G(r) =  {(a-c)(b-c) \over \epsilon \;
c }\; g_{0} \;  U_{\epsilon +1,j} \; ,
 \qquad  F(r) = i \; { a b \over
\epsilon \; c}  \; g_{0} \; U_{\epsilon -1,j}\; ,
\nonumber
\\
\end{eqnarray}

\noindent
quantization rule  $
 \epsilon = 2n + j -1 +3 /2  + \sqrt{m^{2} + 1/  4}$.

\vspace{3mm}

{\bf $(j+1)$-wave }, \qquad \underline{ $j=0,1,2,3, ...$}
\begin{eqnarray}
\vec{A}= e^{-i\epsilon t}
 \; \sqrt{{2j + 1 \over j + 1 }} \;  f (r) \; \vec{J}^{j+1}_{jm}(\theta,\phi) \; ,
 \qquad f(r) = f_{0} \; U_{\epsilon, j+1} \; ,
\nonumber
\\
{1 \over 2} (A^{0} + i A^{5})  =  i\;  G(r) \;  e^{-i(\epsilon
+1)t}  \; Y_{jm} \; , \qquad G(r) =  {j +3/2 \over \epsilon }\;
f_{0} \;  U_{\epsilon +1,j} \; , \nonumber
\\
{1 \over 2} \; ( A^{0} - i A^{5} ) =   F(r) \;  e^{-i(\epsilon
-1)t}  \; Y_{jm} \;, \qquad  F(r) = i \; {j+3/2  \over
\epsilon } \; f_{0} \; U_{\epsilon -1,j}\; ,
\nonumber
\\
\end{eqnarray}

\noindent
quantization rule $\epsilon = 2n + j +1 +3 /2  + \sqrt{m^{2} + 1/  4} \;$.
Degeneration of the energy levels can be  clarified by the table
\begin{eqnarray}
\left.   \begin{array}{rrrr}
     &   \qquad    j-\mbox{type}            &  \qquad   (j-1)-\mbox{type}     &   \qquad    (j+1)-\mbox{type}  \\[1mm]
N=1  &   \qquad    n=0,\; j=1              &  \qquad    n=0,\; j=2          &   \qquad    n=0,\; j=0  \\
N=2  &   \qquad    n=0,\; j=2              &  \qquad    n=0,\; j=3          &   \qquad    n=0,\; j=1  \\
     &   \qquad                            &  \qquad    n=0,\; j=1          &   \qquad                \\
N=3  &   \qquad    n=0,\; j=3              &  \qquad    n=0,\; j=4          &   \qquad    n=0,\; j=2  \\
     &   \qquad    n=1,\; j=1              &  \qquad    n=1,\; j=2          &   \qquad    n=1,\; j=0  \\
N=4  &   \qquad    n=0,\; j=4              &  \qquad    n=0,\; j=5          &   \qquad    n=0,\; j=3  \\
     &   \qquad    n=1,\; j=2              &  \qquad    n=1,\; j=3          &   \qquad    n=1,\; j=1  \\
N=5  &   \qquad    n=0,\; j=5              &  \qquad    n=0,\; j=6          &   \qquad    n=0,\; j=4  \\
     &   \qquad    n=1,\; j=3              &  \qquad    n=1,\; j=4          &   \qquad    n=1,\; j=2  \\
     &   \qquad    n=2,\; j=1              &  \qquad    n=2,\; j=2          &   \qquad    n=2,\; j=0  \\
N=6  &   \qquad    n=0,\; j=6              &  \qquad    n=0,\; j=7          &   \qquad    n=0,\; j=5  \\
     &   \qquad    n=1,\; j=4              &  \qquad    n=1,\; j=5          &   \qquad    n=1,\; j=3  \\
     &   \qquad    n=2,\; j=2              &  \qquad    n=2,\; j=3          &  \qquad    n=2,\; j=1  \\
     &   \qquad                            &  \qquad    n=3,\; j=1          &   \qquad
\end{array} \right.
\nonumber
\end{eqnarray}
\begin{eqnarray}
\left.   \begin{array}{rrrr}
N=7  &   \qquad    n=0,\; j=7              &  \qquad    n=0,\; j=8          &   \qquad    n=0,\; j=6  \\
     &   \qquad    n=1,\; j=5              &  \qquad    n=1,\; j=6          &   \qquad    n=1,\; j=4  \\
     &   \qquad    n=2,\; j=3              &  \qquad    n=2,\; j=4          &  \qquad     n=2,\; j=2  \\
     &   \qquad    n=3,\; j=1              &  \qquad    n=3,\; j=2          &   \qquad    n=3,\; j=0  \\
N=8  &   \qquad    n=0,\; j=8              &  \qquad    n=0,\; j=9          &   \qquad    n=0,\; j=7  \\
     &   \qquad    n=1,\; j=6              &  \qquad    n=1,\; j=7          &   \qquad    n=1,\; j=5  \\
     &   \qquad    n=2,\; j=4              &  \qquad    n=2,\; j=5          &  \qquad     n=2,\; j=3  \\
     &   \qquad    n=3,\; j=2              &  \qquad    n=3,\; j=3          &   \qquad    n=3,\; j=1  \\
     &   \qquad                            &  \qquad    n=4,\; j=1          &  \qquad
     \end{array} \right.
   \nonumber
   \end{eqnarray}

\noindent
where energy level are given by( at $ j=0$,   we have  $\nu = j+1 = 1$)
\begin{eqnarray}
\epsilon = N + {3 \over 2} + \sqrt{m^{2} +1/4}\; , \qquad  N = 2n + \nu \; ,
\nonumber
\\
j=1,2,3, ...\;, \;\;  \nu = j, j-1, j+1 \;.
\label{3.10}
\end{eqnarray}

\section{ Tetrad form of  Maxwell equations in
Riemann -- Silberstein -- Majorana -- Oppenheimer approach
 }

It is well-known that Special Relativity arose  from investigation of
symmetry properties of  the Maxwell equations with respect to
inertial motion of the reference frame: Lorentz \cite{Lorentz-1904}, Poincar\'{e}
\cite{Poincare-1905}, Einstein  \cite{Einstein-1905}. Naturally, it was electromagnetic field that
was the first and principal object for the Special Relativity:
Minkowski \cite{Minkowski-1908}, Silberstein  \cite{Silberstein-1907}, Marcolongo \cite{Marcolongo-1912},
Bateman   \cite{Bateman-1915}.
In  1931 Majorana  \cite{Majorana-1931}   and Oppenheimer   \cite{Oppenheimer-1931}  proposed to
consider classical Maxwell equations as a quantum photon
equations. In this context they introduced 3-vector function
obeying Dirac-like massless wave equation. It turned out that much
earlier in  1907 the same mathematical translation of classical
Maxwell theory was performed by Silberstein \cite{Silberstein-1907}; besides, he
 noted himself that the same approach was used earlier by Riemann
\cite{Weber-1901}. That history was much  forgotten, and  many years this
complex approach to electrodynamics was connected mainly with
Majorana  and  Oppenheimer. Historical justice was rendered by
Bialynicki-Birula  \cite{Bialynicki-Birula-1994},  see also in
 Sipe \cite{Sipe-1995},
 Gersten \cite{Gersten-1997},
 Esposito \cite{Esposito-1997}, Dvoeglazov \cite{Dvoeglazov-1998},
 Ivezic \cite{Ivezic-2006},
 Varlamov \cite{Varlamov-2003},
 Red'kov et al  \cite{Red'kov-Bogush-Tokarevskaya-Spix-2009}.

Below we  use
the complex formalism of Riemann -- Silberstein -- Majorana --
Oppenheimer in Maxwell electrodynamics extended to the case of
arbitrary pseudo-Riemannian space -- time in accordance with the
tetrad recipe of Tetrode -- Weyl -- Fock -- Ivanenko (for more
detail, see  \cite{Bogush-Krylov-Ovsiyuk-Red'kov-2009, Book-2009}).

Maxwell equations in Riemann space can be presented in Riemann -- Silberstein -- Majorana -- Oppenheimer
basis as one matrix equation
\begin{eqnarray}
\alpha^{c} \; ( \; e_{(c)}^{\rho} \partial_{\rho} + {1 \over 2}
j^{ab} \gamma_{abc} \; ) \; \Psi = J(x)\; , \nonumber
\\
 \alpha^{0} = -i I\; ,
\qquad \Psi = \left | \begin{array}{c} 0 \\ {\bf E} + i c{\bf B}
\end{array} \right | \; , \qquad J
= {1 \over \epsilon_{0}} \; \left | \begin{array}{c} \rho \\
i{\bf j}
\end{array} \right |
\noindent
\label{3.1'}
\end{eqnarray}

\noindent  or
\begin{eqnarray}
-i  (  e_{(0)}^{\rho} \partial_{\rho} + {1 \over 2} j^{ab}
\gamma_{ab0}  )\Psi + \alpha^{k}  (  e_{(k)}^{\rho}
\partial_{\rho} + {1 \over 2} j^{ab} \gamma_{abk}  )\Psi =
J(x)\; . \label{3.2'}
\end{eqnarray}

\noindent Allowing for identities
\begin{eqnarray}
{1 \over 2} j^{ab} \gamma_{ab0} = [ s_{1} ( \gamma_{230} +i
\gamma_{010} ) + s_{2} ( \gamma_{310} +i \gamma_{020}) + s_{3} (
\gamma_{120} +i \gamma_{030} ) ] \; , \nonumber
\\
{1 \over 2} j^{ab} \gamma_{abk} = [ s_{1} ( \gamma_{23k} +i
\gamma_{01k} ) + s_{2} ( \gamma_{31k} +i \gamma_{02k}) + s_{3} (
\gamma_{12k} +i \gamma_{03k} ) ] \; ,
\nonumber \label{3.3'}
\end{eqnarray}

\noindent and using the notation
\begin{eqnarray}
e_{(0)}^{\rho} \partial_{\rho} = \partial_{(0)} \; , \qquad
e_{(k)}^{\rho} \partial_{\rho} = \partial_{(k)} \; , \qquad
a =0,1,2,3 \; , \nonumber
\\
( \gamma_{01a}, \gamma_{02a} , \gamma_{03a} ) = {\bf v}_{a} \; ,
\qquad ( \gamma_{23a}, \gamma_{31a} , \gamma_{12a} ) = {\bf p}_{a}
\; , \label{3.4'}
\end{eqnarray}

\noindent eq. (\ref{3.2'})  in absence of sources reduces to
\begin{eqnarray}
-i  [ \;  \partial_{(0)} + {\bf s} ({\bf p}_{0} +i{\bf v}_{0}
) \; ] \Psi    + \alpha^{k}\;   [ \; \partial_{(k)} + {\bf s} ({\bf
p}_{k} +i{\bf v}_{k}  )\;  ] \;   \Psi =  0 \; ,
\label{3.5'}
\end{eqnarray}

\noindent where
\begin{eqnarray}
s_{1}=  \left | \begin{array}{cc} 0 & 0 \\0 & \tau_{1} \end{array}
\right |,\;
 s_{2} =   \left | \begin{array}{cc}
0 & 0 \\0 & \tau_{1} \end{array} \right |  , \;
s_{3} =  \left | \begin{array}{cc}
0 & 0 \\0 & \tau_{1} \end{array} \right |  ,
\nonumber
\\
\tau_{1} = \left | \begin{array}{rrr}
 0 & 0 & 0 \\
0 & 0 & -1 \\
 0 & 1 & 0 \\
\end{array} \right | ,
 \tau_{2} = \left | \begin{array}{rrr}
0 & 0 & 1 \\
 0 & 0 & 0 \\
 -1 & 0 & 0 \\
\end{array} \right | , \;
\tau_{3} = \left | \begin{array}{rrr}
 0 & -1 & 0 \\
 1 & 0 & 0 \\
 0 & 0 & 0
\end{array} \right |  .
\label{tau}
\end{eqnarray}

With the use of spherical tetrad
in the anti de Sitter space
the main equation (\ref{3.5'}) takes the form
\begin{eqnarray}
 [   -  { i  \partial_{t} \over \sqrt{ \Phi }}  + \sqrt{\Phi}  (
 \alpha^{3}  \partial_{r}   + {  \alpha^{1}  s_{2}  - \alpha^{2}  s_{1}   \over r} +
    { r \over   \Phi }  s_{3}  )
   +  {1 \over r }   \Sigma_{\theta, \phi}       ]
 \left | \begin{array}{c}
0 \\ \psi
\end{array} \right | = 0 \; ,
\nonumber
\\
\Sigma_{\theta, \phi} =
  { \alpha^{1}  \over r}\partial_{\theta} +
 \alpha^{2} \;  { \partial_{\phi} +   s_{3}   \cos \theta  \over  \sin \theta } \;   .
\label{4.11'}
\end{eqnarray}

It is convenient to have the spin matrix  $s_{3}$ as  diagonal, which is reached by simple
linear transformation to the known cyclic basis
\begin{eqnarray}
 \Psi ' = U_{4} \Psi  \; , \qquad
\; U_{4}  = \left | \begin{array}{cc} 1 & 0 \\
0 &   U
\end{array} \right |  ,
\nonumber
\\
U = \left |
 \begin{array}{ccc}
- 1 /\sqrt{2}  &  i /  \sqrt{2}  &  0  \\
0  &  0  &  1  \\
1 / \sqrt{2}  &  i  / \sqrt{2}  &  0
\end{array} \right |  , \;
U^{-1}  = \left |
 \begin{array}{ccc}
- 1 /\sqrt{2}  &  0  & 1 /  \sqrt{2}    \\
-i / \sqrt{2}  &  0  &  -i / \sqrt{2}   \\
0    &  1    &  0
\end{array} \right |  .
\label{cyclic}
\end{eqnarray}

\noindent  so that
\begin{eqnarray}
U  \tau_{1} U^{-1} = {1 \over \sqrt{2}} \left |
\begin{array}{ccc}
0  &  -i   &  0  \\
-i  &  0  &  -i  \\
0  &  -i  &  0
\end{array} \right | =  \tau'_{1} \; ,
\qquad j^{'23} = s_{1}' = \left | \begin{array}{cc}
0 & 0 \\
0 & \tau'_{1}
\end{array} \right | \; ,
\nonumber
\\
U  \tau_{2} U^{-1} = {1 \over \sqrt{2}} \left |
\begin{array}{ccc}
0  &  -1  &  0  \\
1  & 0  &  -1  \\
0  &  1  &  0
\end{array} \right | =  \tau'_{2}\; , \qquad  j^{'31} = s_{2}' = \left | \begin{array}{cc}
0 & 0 \\
0 & \tau'_{2}
\end{array} \right | \;
, \nonumber
\end{eqnarray}
\begin{eqnarray}
U  \tau_{3} U^{-1} = - i \; \left | \begin{array}{rrr}
+1  &  0  &  0  \\
0  &  0  &  0   \\
0  &  0  &  -1
\end{array} \right | = \tau'_{3} \; \qquad  j^{'12} = s_{3}' = \left | \begin{array}{cc}
0 & 0 \\
0 & \tau'_{3}
\end{array} \right | \;
. \; \nonumber
\\
 \alpha^{'1} = {1 \over \sqrt{2}}
\left | \begin{array}{rrrr}
0 & - 1 & 0 & 1 \\
1 & 0 & -i & 0 \\
0 & -i & 0 & - i \\
-1 & 0 & -i & 0
\end{array} \right |  ,
\alpha^{'2} = {1 \over \sqrt{2}} \left | \begin{array}{rrrr}
0 & - i & 0 & -i \\
-i & 0 & -1 & 0 \\
0 & 1 & 0 &  -1 \\
-i & 0 & 1 & 0
\end{array} \right |  ,
  \alpha^{'3}  =  \left
| \begin{array}{rrrr}
0  & 0  & 1 & 0 \\
0  & -i & 0 & 0 \\
-1 &  0 & 0 & 0 \\
0  &  0 & 0 & +i
\end{array} \right |  .
\nonumber
\end{eqnarray}

Eq.  (\ref{4.11'})  becomes
\begin{eqnarray}
 [   -  { i  \partial_{t} \over \sqrt{ \Phi }}  + \sqrt{\Phi}  (
 \alpha^{'3}  \partial_{r}   + {  \alpha^{'1}  s_{'2}  - \alpha^{'2}  s'_{1}   \over r} +
    { r \over   \Phi } s'_{3}  )
   +  {1 \over r }   \Sigma'_{\theta, \phi}       ]
 \left | \begin{array}{c}
0 \\ \psi'
\end{array} \right | = 0 \; ,
\nonumber
\\
\Sigma'_{\theta, \phi} =
  { \alpha^{'1}  \over r}\partial_{\theta} +
 \alpha^{'2} \;  { \partial_{\phi} +   s'_{3}   \cos \theta  \over  \sin \theta } \;   .
\label{4.11''}
\end{eqnarray}

\section{Separating the variables and Wigner functions}

\hspace{5mm} Let us diaginalize operators  ${\bf J}^{2}, J^{3}$
-- corresponding substitution for $\psi$ is
\begin{eqnarray}
\psi  = e^{-i \omega t} \left | \begin{array}{l}
0 \\
f_{1}(r) D_{-1 }
\\
f_{2}(r) D_{0 }  \\
f_{3}(r) D_{+1 }
\end{array} \right |
\label{5.1}
\end{eqnarray}

\noindent where the shorted notation for Wigner $D$-functions
$ D_{\sigma} = D^{j}_{-m, \sigma} ( \phi , \theta, 0)\;,
\; \sigma = -1, 0, +1 $;
$j,m$  determine total angular moment.
With the use  the following recursive relations  \cite{Varshalovich-Moskalev-Hersonskiy-1975}
\begin{eqnarray}
\partial_{\theta}   D_{-1} =   {1 \over 2}  (
a   D_{-2} -    \nu   D_{0}  ) \; , \; \frac {m -
\cos{\theta}}{\sin {\theta}}  D_{-1} = {1 \over 2} (  a
D_{-2} + \nu  D_{0}  ) \; , \nonumber
\\
\partial_{\theta}   D_{0} = {1 \over 2}  (
\nu   D_{-1} - \nu   D_{+1}  ) \; , \; \frac {m}{\sin
{\theta}}  D_{0} = {1 \over 2}  (    \nu  D_{-1} +  \nu
D_{+1}  ) \; , \nonumber
\\
\partial_{\theta}  D_{+1} =
{1 \over 2}  (  \nu   D_{0} - a   D_{+2}  ) \; , \;
\frac {m + \cos{\theta}}{\sin {\theta}}   D_{+1} ={1 \over 2} \;
(
 \nu  D_{0} + \ a   D_{+2}  ) \;  ,
\nonumber
\\
\nu =  \sqrt{j(j+1)} \; , \qquad  a = \sqrt{(j-1)(j+2)} \; .
\label{5.2'}
\end{eqnarray}

\noindent we get
 (the factor  $e^{-i \omega
t} $ is omitted)
\begin{eqnarray}
 \Sigma'_{\theta \phi} \Psi ' =
 { \nu  \over \sqrt {2}} \;
 \left | \begin{array}{r}
 (f_{1} +f_{3}) D_{0} \\
 -i \;  f_{2} D_{-1}
 \\
 i \;  (f_{1} -f_{3}) D_{0}
 \\
 + i \;  f_{2} D_{+1}
 \end{array} \right  |
\label{5.3'}
\end{eqnarray}

Turning back to Maxwell equation  (\ref{4.11''}), after simple calculation we arrive at
the radial system
\begin{eqnarray}
(1) \qquad   \sqrt{ \Phi }\; ( {d \over d r} +  {2   \over r }
)\;f_{2} + {1 \over r } \;
 { \nu  \over \sqrt {2}} \;  (f_{1} +f_{3})= 0\; ,
\nonumber
\\
(2) \qquad   ( - {\omega \over \sqrt{ \Phi }}   - i\; \sqrt{
\Phi }\; {d \over dr } - i {\sqrt{ \Phi }  \over r }  - i{r\over
\sqrt{ \Phi }}  ) \;f_{1} -
 {i \over r } \;  { \nu  \over \sqrt {2}} \; f_{2} = 0 \; ,
\nonumber
\\
(3) \qquad  \qquad - {\omega \over \sqrt{ \Phi }} f_{2} + {i \over
r } \;
 { \nu  \over \sqrt {2}} (f_{1} -f_{3}) = 0 \; ,
\nonumber
\\
(4) \qquad  (- {\omega \over \sqrt{ \Phi }} + i \; \sqrt{
\Phi }\;{d \over dr }  + i {\sqrt{ \Phi }  \over r } +i{r\over
\sqrt{ \Phi }} ) \;f_{3} + {i \over r } \;  { \nu  \over
\sqrt {2}} \; f_{2} = 0 \; . \label{5.5'}
\end{eqnarray}

Combining equations  (2) and  (4),  instead of  (\ref{5.5'}) we get
\begin{eqnarray}
 (2) + (4)  , \qquad
  - {\omega \over \sqrt{ \Phi }} (f_{1} + f_{3})   -
i   ( \sqrt{ \Phi } {d \over dr } + {\sqrt{ \Phi } \over
r } + {r\over \sqrt{ \Phi }}   )   (f_{1} - f_{3}) = 0
, \nonumber
\\
(2) -(4)  , \;
  - {\omega \over \sqrt{ \Phi }}   (f_{1} - f_{3})  - i (  \sqrt{ \Phi
} {d \over dr }  + {\sqrt{ \Phi }  \over r } + {r\over \sqrt{
\Phi }}   ) (f_{1} + f_{3})  -
 {2i \over r }   { \nu  \over \sqrt {2}}  f_{2}  = 0  ,
\nonumber
\\
(3) \qquad  - {\omega \over \sqrt{ \Phi }} f_{2} + {i \over
r }  { \nu  \over \sqrt {2}} (f_{1} -f_{3}) = 0  ,
\nonumber
\\
(1) \qquad   \sqrt{ \Phi } ( {d \over d r} +  {2   \over r }
)f_{2} + {1 \over r }
 { \nu  \over \sqrt {2}}   (f_{1} +f_{3})= 0.
\nonumber
\label{3.9}
\end{eqnarray}

\noindent
It is easily verified that equation  (1) is an identity when allowing for remaining ones.
So independent equations are
\begin{eqnarray}
- {\omega \over \sqrt{ \Phi }} f_{2} + {i \over r } \;
 { \nu  \over \sqrt {2}} (f_{1} -f_{3}) = 0 \; ,
\nonumber
\\
  - {\omega \over \sqrt{ \Phi }} (f_{1} + f_{3})   -
i\;   ( \sqrt{ \Phi }\; {d \over dr } + {\sqrt{ \Phi } \over
r } + {r\over \sqrt{ \Phi }}   )   (f_{1} - f_{3}) = 0 \;
, \nonumber
\\
  - {\omega \over \sqrt{ \Phi }}   (f_{1} - f_{3})  - i\; (  \sqrt{ \Phi
} {d \over dr }  + {\sqrt{ \Phi }  \over r } + {r\over \sqrt{
\Phi }}  )  (f_{1} + f_{3})  -
 {2i \over r }   { \nu  \over \sqrt {2}}  f_{2}  = 0 \; .
\label{3.10'}
\end{eqnarray}

Let us introduce new functions:
$
f = ( f_{1} + f_{3} ) /  \sqrt{2}   \; , \;  g = ( f_{1}-
f_{3}  ) / \sqrt{2}$,
 then eqs.  (\ref{3.10'}) look
\begin{eqnarray}
 f_{2} = {i  \nu \over \omega  }   { \sqrt{ \Phi }  \over r }  g  = 0 \; ,
\;\; - {\omega \over  \Phi }  f   -
i    (  {d \over dr } + {1  \over r } + {r\over \Phi }
 )   g = 0 \; , \nonumber
\\
  - {\omega^{2} \over \Phi  }   g  - i \omega   (   {d \over dr }  + { 1  \over r } + {r\over
\Phi }  ) \; f  +
 {\nu^{2}  \over r^{2} } \;    \; g   = 0 \; ,
\label{3.12'}
\end{eqnarray}

The system   (\ref{3.12'})  is simplified by substitutions
\begin{eqnarray}
g = {1\over r \sqrt {1+r^{2}}}\; G (r) \; , \; f = {1\over r \sqrt
{1+r^{2}}}\;  F (r),
\nonumber
\\
 f_{2} = {i  \nu \over \sqrt{2} \omega  } \;
 {1\over r^{2}  }\; G (r)   = 0 \; , \qquad
 i  \omega \;  F  = \Phi {d \over dr } G \; ,
 \;\nonumber
 \\
  i \omega \;   {d \over dr } \; F  +   {\omega^{2} \over \Phi  }   \; G   - {\nu^{2} \over r^{2}} \; G   = 0 \; ,
\label{3.13'}
\end{eqnarray}

\noindent So we have arrived at a   single  differential equation
for  $G (r)$:
\begin{eqnarray}
(1+r^{2})\;{d^{2}G\over d r^{2}}+2r{d G \over
dr}+\left({\omega^{2}\over 1+r^{2}}-{\nu^{2}\over
r^{2}}\right)G=0. \label{3.14}
\end{eqnarray}

\noindent
In  (\ref{3.14}) let us introduce a new variable
$z= -r^{2}$, which results in
\begin{eqnarray}
4z(1-z){d^{2}G\over dz^{2}}+2(1-3z){dG\over dz}-\left(
{\omega^{2}\over 1-z}+{\nu^{2}\over z}\right)G=0, \label{3.15'}
\end{eqnarray}

\noindent with the use of substitution
$
G = z^{a} (1-z)^{b} F (z).
$ eq.  (\ref{3.15'}) gives
\begin{eqnarray}
4z(1-z){d^{2}F\over
dz^{2}}+4\left[2a+{1\over2}-(2a+2b+{3\over2})z\right]{dF\over dz}+
\nonumber
\\
+\left[{4a^{2}-2a-\nu^{2}\over z} +{4b^{2}-\omega^{2}\over
1-z}-4(a+b)(a+b+{1\over2})\right]F=0. \label{3.16'}
\end{eqnarray}

With  requirements
\begin{eqnarray}
4a^{2}-2a-\nu^{2} = 0 \; \; \Longrightarrow \; \nonumber
\\
a={1\over 4} \pm \; {1 \over 4} \sqrt{1+4 \nu^{2}}  =
 {1 \over 4} \pm {1\over 2}(j +{1 \over 2})  = -{j \over 2} , + {j+1 \over 2} \; ,
\nonumber
\\
4b^{2}-\omega^{2} = 0 \; \; \Longrightarrow \; \; b = \pm \; {
\omega  \over 2} \; , \; \omega > 0 \;  ;
 \label{3.17'}
\end{eqnarray}

\noindent to have solutions vanishing ar $r=0$ one musr take positive values
 $a =  (j+1)/2 $, eq. (\ref{3.16'}) take the form

\begin{eqnarray}
z(1-z) {d^{2} F\over
dz^{2}} + [ \; 2a+{1\over2}-(2a+2b+{3\over2})z \; ] \; {dF\over
dz}
- (a+b)(a+b+{1\over2}) \; F=0 \; , \label{3.18}
\end{eqnarray}

\noindent which is an equation of hypergeometric type
\begin{eqnarray}
\gamma = 2a +{1\over2} \; , \qquad \alpha + \beta = 2a +
2b+{1\over2} \;, \qquad \alpha \beta = (a+b)(a+b+ {1\over2})  \; ,
\nonumber
\end{eqnarray}

\noindent that is
\begin{eqnarray}
\alpha = a+b \;, \qquad \beta = a+b+ {1\over2} \;, \qquad \gamma =
2a +{1\over2} \; .
 \label{3.19}
\end{eqnarray}

\noindent To have polynomials one must take negative value for
 $b= - \omega /2$. So, parameters are
 \begin{eqnarray}
\alpha  =  {j +1 \over 2} - {\omega \over  2} \; , \qquad \beta  =
{j +1 \over 2} - {\omega \over 2} + {1 \over 2}  \; ,
 \label{3.20}
\end{eqnarray}

\noindent

\noindent
and quantization  is given by
\footnote{ There exists symmetric variant $ \beta = - n \;
\Longrightarrow \;   \omega_{n,j+1} = 2n + (j +1 ) + 1 $. }:
\begin{eqnarray}
\alpha= - n  \; , \qquad  \omega_{n,j} = 2n + j +1n  \; , \qquad
(n= 0, 1, 2, ...) \;;
 \label{3.21'}
\end{eqnarray}

\noindent or in usual units
$
\omega = (c /  \rho ) \; ( 2n +  j + 1  )\; ;
$ $\rho$ is a curvature radius..

\section{ Acknowledgment }

 Authors are  grateful  to  participants of
seminar of Laboratory of Physics of Fundamental Interaction,
 National Academy of Sciences of Belarus, for stimulating discussion and  advice.


\begin{thebibliography}{xxx}




\bibitem{Dirac-1935}
  Dirac, P.A.M.:
The electron wave equation in the de Sitter space. Ann. Math. {\bf 36},  657--669 (1935)


\bibitem{Dirac-1936}
  Dirac, P.A.M.:
 Wave equations in conformal space.
 Ann. of Math.  {\bf 37},  429--442 (1936)



\bibitem{Schrodinger-1939}
 Schr\"{o}dinger, E.:
The proper vibrations of the expanding universe. Physica. {\bf 6},  899--912 (1939)



\bibitem{Schrodinger-1940}
 Schr\"{o}dinger,   E.:
 General theory of relativity and wave mechanics. Wiss. en Natuurkund. {\bf 10},  2--9 (1940)








\bibitem{Lubanski-Rosenfeld-1942}
Lubanski,  J.K.,  Rosenfeld,  L.:
 Sur la representation des champs mesoniques dans l'\'espace \`a sinq
dimension.  Physica.  {\bf  9},  117 (1942)


\bibitem{Goto-1951}
Goto, K.:
 Wave equations in de Sitter space.
Progr. Theor. Phys.  {\bf  6}, 1013--1014 (1951)


\bibitem{Ikeda-1953}
 Ikeda, M.:
On a five-dimensional representation of the
electromagnetic and electron field equations in a curved
space-time.  Progr. Theor. Phys.  {\bf 10},  483--498 (1953)



\bibitem{Nachtmann}
Nachtmann, O.:
Quantum theory in de-Sitter space. Commun. Math. Phys. {\bf 6}, 1--16 (1967)



\bibitem{Chernikov-Tagirov}
Chernikov, N.A., Tagirov, E.A.:
Quantum theory of scalar field in de Sitter space-time. Ann.
Inst. Henri Poincare {\bf IX}, 109--141 (1968)





\bibitem{Geheniau-Schomblond}
Geheniau, J., Schomblond, Ch.:
Functions de Green dans l'Univers de de Sitter. Bull. Cl. Sci.,
V. Ser., Acad. R. Belg. {\bf 54}, 1147--1157 (1968)



\bibitem{Borner-Durr-1969}
 B\"{o}rner,  G.,    D\"{u}rr, H.P.:
 Classical and quantum theory in de Sitter space.
Nuovo Cim. A. {\bf  64}, 669--713 (1969)

\bibitem{Tugov-1969}
I.I. Tugov.
Conformal covariance and invariant formulation of scalar wave equations
Ann. Inst. Henri Poincar\'e.   A.  {\bf 11}, 207--220  (1969)


\bibitem{Fushchych-Krivsky-1969}
Fushchych, W.L.,  Krivsky, I.Yu.:
On  representations of the
inhomogeneous de Sitter group and equations in five-dimensional
Minkowski space.   Nucl. Phys. B.   {\bf 14},  573--585 (1969)



\bibitem{Borner-Durr}
B\"{o}rner G., D\"{u}rr,  H.P.:
Classical and Quantum Fields in de Sitter space.
 Nuovo Cim. {\bf  LXIV}, 669    (1969)




\bibitem{Chevalier-1970}
Chevalier M.:
L'\'{e}quation de Kirchjgoff g\'{e}n\'{e}ralis\'{e}e.
Ann. Inst. Henri Poincar\'e.   A.  {\bf 12}, 71--115  (1970)





\bibitem{Castagnino-1970}
Castagnino, M.:
Champs de spin entier dans l'espace-temps De Sitter.
Ann. Inst. Henri Poincar\'e.   A.  {\bf 13}, 263--270  (1970)


\bibitem{Vidal-1970}
 Vidal, A.:
On the consistency of wave equations in de Sitter space.
 Notas Fis. {\bf 16},   8 (1970)



\bibitem{Adler-1972}
 Adler,   S.L.:
 Massless, euclidean quantum electrodinamics on the
5-dimensional unit hypersphere. Phys. Rev. D. {\bf  6},  3445--3461 (1972)



\bibitem{Castagnino-1972}
Castagnino, M.:
Champs spinoriels en Relativit\'e g\'en\'erale; le cas particulier
de l\'espace-temps de De Sitter et les \'equations  d'ond pour les spins \'el\'eves
Ann. Inst. Yenri Poincar\'e. A {\bf 16}. 293--341 (1972)


\bibitem{Schnirman-Oliveira-1972}
Schnirman, E., Oliveira, C.G.:
Conformal invariance of the equations of motion in curved spaces.
Ann. Inst. Henri Poincar\'e. A {\bf 17},  379--397  (1972)


\bibitem{Tagirov-1973}
Tagirov, E.A.:
Consequences of field quantization in de Sitter type cosmological models.
Ann. Phys. {\bf 76}, 561--579 (1973)



\bibitem{Riordan-1974}
 Riordan, F.:
Solutions of the Dirac equation in finite de Sitter space.
Nuovo Cim.  B.   {\bf 20}, 309--325 (1974)



\bibitem{Pestov-Chernikov-Shavoxina-1975}
 Pestov,  F.B., Chernikov, N.A., Shavoxina, N.S.:
 Electrodynamical equations in spherical world. Teor. Mat. Fiz.
  {\bf  25},  327--334 (1975)



\bibitem{Candelas-Raine-1975}
Candelas, P.,  Raine, D.J.:
General-relativistic quantum field theory: an exactly soluble model.
 Phys. Rev. D.  {\bf 12}, 965--974 (1975)


\bibitem{Schomblond-Spindel-1976}
Schomblond, C., Spindel P.:
Propagateurs des champs spinoriels et vectoriels dans l'univers de de Sitter.
 Acad. Roy. Belgium.  Class Sci LXII, 5eme serie (1976) 124



\bibitem{Schomblond-Spindel-1976'}
Schomblond, Ch., Spindel, P.:
Conditions d'unicite pour le propagateur $\Delta^{1}(x,y)$ du champ
scalaire dans l'univers de de Sitter.
Ann. Inst. Henri Poincare. {\bf  XXV}, 67--78 (1976)


\bibitem{Dowker-Critchley-1976}
Dowker, J.S., Critchley, R.:
Scalar effective Lagrangian in de Sitter space.
Phys. Rev. D {\bf 13}, 224--234 (1976)


\bibitem{Avis-Isham-Storey-1978}
 Avis, S.J.,  Isham,  C.J.,  Storey,  D.:
  Quantum Field Theory In Anti-de Sitter Space-Time.
  Phys. Rev. D {\bf 18},   3565 (1978)


\bibitem{Brugarino-1980}
Brugarino, T.
De Sitter-invariant field equations.
Ann. Inst. Henri Poincar\'e. A
{\bf 32},  277--282 (1980)



\bibitem{Fang-Fronsdal-1980}
 Fang, J. Fronsdal,  C.:
Massless, half-integer-spon fields in de Sitter space
 Phys. Rev. D {\bf 22}, 1361--1367  (1980)




\bibitem{Angelopoulos-Flato-Fronsdal-Sternheimer-1981}
 Angelopoulos, E.,  Flato,  M.,  Fronsdal,  C.,  Sternheimer  D.:
Massless Particles, Conformal Group, and De Sitter Universe.
 Phys. Rev. D {\bf 23},  1278--1289  (1981)


\bibitem{Burges-1984}
Burges, C.J.C:
The de Sitter vacuum. Nucl. Phys. B {\bf 247}, 533--543 (1984)




\bibitem{Deser-Nepomechie-1984}
 Deser, S.,   Nepomechie, R.I.:
 Gauge Invariance Versus Masslessness in de Sitter
Space. Ann. Phys. {\bf 154}   396--420   (1984)



\bibitem{Dullemond-Beveren-1985}
Dullemond, C, van Beveren, E.:
Scalar field propagators in anti-de Sitter spacetime.
 J. Math. Phys. {\bf 26}, 2050--2058 (1985)



\bibitem{Gazeau-1985}
Gazeau, J.P.:
Gauge fixing and Gupta-Bleuler triplet in de Sitter QED.
J. Math. Phys. {\bf 26}, 1847--1854 (1985)



 \bibitem{Allen-1985}
 Allen, B.:
 Vacuum states in de Sitter space.
  Phys. Rev. D {\bf 32}, 3136--3149 (1985)




\bibitem{Fefferman-Graham-1985}
 Fefferman, C.,  Graham, C.R.:
  Conformal invariants, in: \'{E}lie
Cartan et les Mathematiques d'Aujourd'hui, Ast\'{e}risque,
numero hors serie, Soc. Math. France, Paris, 95--116 (1985)










 \bibitem{Flato-Fronsdal-Gazeau-1986}
  Flato, M.,  Fronsdal,   C.,  Gazeau, J.P.:
 Masslessness and light-cone propagation in 3+2 de Sitter and 2+1 Minkowski spaces
 Phys. Rev. D {\bf 33}, 415--420 (1986)



\bibitem{Allen-Jacobson-1986}
Allen, B., Jacobson, T.:
 Vector two-point functions in maximally
symmetric space. Commun. Math. Phys. 103, 669--692 (1986)




\bibitem{Allen-Folacci-1987}
 Allen, B., Folacci,  A.:
 The Massless Minimally Coupled Scalar Field In
De Sitter Space.
Phys. Rev. D {\bf 35}, 3771--3778 (1987)




\bibitem{Sanchez-1987}
 S\'{a}nchez, N.:
Quantum field theory and elliptic interpretation
of de Sitter space-time.
Nucl. Phys. B. {\bf 294}, 1111--1137 (1987)



\bibitem{Pathinayake-Vilenkin-Allen-1988}
  Pathinayake, C.,  Vilenkin,   A.,  Allen,  B.:
 Massless scalar and antisymmetric tensor fields in de Sitter space.
 Phys. Rev. D {\bf 37}, 2872--2877 (1988)



\bibitem{Gazeau-Hans-1988}
Gazeau, J-P., Hans,  M.:
Integral-spin fields on (3+2)-de Sitter space.
 J. Math. Phys. {\bf  29}, 2533--2552   (1988)



\bibitem{Bros-Gazeau-Moschella-1994}
Bros J., Gazeau, J.P, Moschella, U.:
Quantum Field Theory
in the de Sitter Universe.
 Phys. Rev. Lett. {\bf  73}, 1746   (1994)



\bibitem{Takook-1997}
Takook, M.V.
 Th\`{e}se de l'universit\'{e} Paris VI, (1997)  Th\'{e}orie quantique des champs pour des
syst\`{e}mes \'{e}l\'{e}mentaires "massifs" et de "masse nulle" sur l'espace-temps de de Sitter.



\bibitem{Pol'shin-1998(1)}
Pol'shin, S.A.:
Group Theoretical Examination of the Relativistic Wave
Equations on Curved Spaces. I. Basic Principles.
arXiv:gr-qc/9803091




\bibitem{Pol'shin-1998(2)}
Pol'shin, S.A.:
Group Theoretical Examination of the Relativistic Wave
Equations on Curved Spaces. II.  De Sitter and Anti-de Sitter Spaces.
arXiv:gr-qc/9803092



\bibitem{Pol'shin-1998(3)}
Pol'shin, S.A.:
Group Theoretical Examination of the Relativistic Wave
Equations on Curved Spaces. III. Real reducible spaces.
arXiv:gr-qc/9809011







\bibitem{Gazeau-Takook-2000}
  Gazeau, J-P.,  Takook,  M.V.:
"Massive" vector field in de Sitter space
J. Math.Phys. {\bf 41},  5920--5933  (2000)



\bibitem{Takook-2000}
  Takook, M.V.:
 Spin 1/2 Field Theory in the de Sitter space-time. arXiv:gr-qc/0005077











\bibitem{Deser-Waldron-2001}
Deser, S.,   Waldron, A.:
Partial masslessness of higher spins in (A)dS.
Nucl. Phys. B {\bf 607}, 577--604    (2001) [hep-th/0103198]


\bibitem{Deser-Waldron-2001'}
Deser, S.,   Waldron, A.:
Null propagation of partially massless higher
spins in (A)dS and cosmological constant speculations.
Phys. Lett. B {\bf 513}, 137 (2001) [hep-th/0105181]









\bibitem{Spradlin-Strominger-Volovich-2001}
 Spradlin, M.,  Strominger,  A.,   Volovich, A.:
 Les Houches lectures on de Sitter space. (29 pages)   hep-th/0110007




\bibitem{Cai-Myung-Zhang-2002}
 Cai, R.G.,  Myung,  Y.S.,  Zhang,  Y.Z.:
  Check of the mass bound
conjecture in de Sitter space, Phys. Rev. D {\bf 65},  084019 (2002),
[hep-th/0110234]



\bibitem{Garidi-Huguet-Renaud-2003}
 Garidi,  T.,  Huguet, E.,  Renaud,  J.:
 De Sitter Waves and the Zero Curvature Limit Comments.
  Phys. Rev. D {\bf 67}, 124028  [5 pages] (2003)









\bibitem{Rouhani-Takook-2005}
 Rouhani, S.,  Takook,  M.V.:
 Abelian Gauge Theory in de Sitter Space
 Mod. Phys. Lett. A {\bf 20},  2387--2396  (2005)




\bibitem{Behroozi-Rouhani-Takook-Tanhayi-2006}
Behroozi, S.,  Rouhani, S., Takook,  M.V.,   Tanhayi, M.R.:
Conformally invariant wave-equations and massless fields in de Sitter spacetime.
 Phys.Rev. D {\bf 74}, 124014 (2006) [arXiv:gr-qc/0512105]






\bibitem{Huguet-Queva-Renaud-2006}
 Huguet, E.,  Queva, J.,  Renaud,  J.:
Conformally related massless fields in dS, AdS and Minkowski spaces
Phys. Rev. D {\bf 73}, 084025  [7 pages] (2006)



\bibitem{Garidi-Gazeau-Rouhani-Takook-2008}
Garidi, T.,  Gazeau, J.P., Rouhani,  S., Takook,  M.V.:
"Massless vector field in de Sitter Universe.
 J. Math. Phys. {\bf 49}, 032501 (2008) [arXiv:gr-qc/0608004]



\bibitem{Huguet-Queva-Renaud-2008(2)}
 Huguet, E.,  Queva, J.,  Renaud,  J.:
Revisiting the conformal invariance of the scalar field:
 From Minkowski space to de Sitter space.
Phys. Rev. D {\bf 77}, 044025  [4 pages] (2008)



\bibitem{Dehghani-Rouhani-Takook-Tanhayi-2008}
  Dehghani, M.,  Rouhani,  S.,  Takook, M.V., Tanhayi,  M.R.:
 Conformally Invariant "Massless" Spin-2 Field in de Sitter Universe.
 Phys. Rev.D {\bf 77}, 064028 (2008)




\bibitem{Moradi-Rouhani-Takook-2008}
 Moradi,  S.,  Rouhani,   S.,  Takook, M.V.:
 Discrete Symmetries for Spinor Field in de Sitter Space
Phys. Lett. B {\bf 613} 74--82 (2005)




\bibitem{Faci-Huguet-Queva-Renaud-2009}
 Faci, S.,  Huguet,   E.,  Queva,    J.,  and  Renaud, J.:
Conformally covariant quantization of the Maxwell field in de Sitter space.
Phys. Rev. D {\bf 80}, 124005  [13 pages] (2009)




\bibitem{Hawking-1971}
Hawking, S.W.:
 Gravitational radiation from colliding black holes. Phys. Rev. Lett. {\bf 26},
 1344--1346 (1971)

\bibitem{Hawking-1974}
Hawking, S.W.:
Black hole explositions?
 Nature. {\bf  248}, no 5443,  30--31 (1974)



\bibitem{Hawking-1975}
Hawking, S.W.:
Particle creation by black  holes.
 Commun. Math. Phys. {\bf  43},  199--220 (1975)






\bibitem{Lohiya-Panchapakesan-1978}
Lohiya, D., Panchapakesan, N.:
Massless scalar field in a de Sitter universe  and its thermal flux.
 J. Phys. A. {\bf  11},  1963--1968 (1978)




\bibitem{Lohiya-Panchapakesan-1979}
Lohiya, D., Panchapakesan, N.:
Particle emission in the de Sitter universe for
massless fields with spin.
J. Phys. A.  {\bf  12},  533--539  (1979)








\bibitem{Khanal-Panchapakesan-1981(1)}
Khanal, U., Panchapakesan, N.:
Perturbation  of the de Sitter-Schwarzchild universe with massless fields.
 Phys. Rev. D.  {\bf  24},    829--834 (1981)




\bibitem{Khanal-Panchapakesan-1981(2)}
Khanal, U., Panchapakesan,  N.:
Production of massless particles in the de Sitter-Schwarzschild universe.
 Phys. Rev. D  {\bf 24}, 835--838 (1981)


\bibitem{Khanal-1983}
 Khanal, U.:
Rotating black hole in asymptotic de Sitter space: Perturbation of the space-time with spin fields
 Phys. Rev. D  {\bf  28}, 1291--1297 (1983)


\bibitem{Hawking-Page-1983}
Hawking, S.,  Page,  D.:
Thermodynamics Of Black Holes In Anti-de Sitter Space.
Commun. Math. Phys. {\bf 87}   577--588 (1983)


\bibitem{Khanal-1985}
 Khanal, U.:
Further investigations of the Kerr-de Sitter space.
 Phys. Rev. D  {\bf  32}, 879--883 (1985)




\bibitem{Otchik-1985}
Otchik, V.S.:
On the Hawking radiation of spin 1/2 particles
in the de Sitter space-time.
 Class. Quantum Crav.  {\bf  2},  539--543 (1985)



\bibitem{Motolla-1985}
Motolla, F. Particle creation in de Sitter space.
 Phys. Rev. D  {\bf  31},  754--766 (1985)





\bibitem{Bogush-Otchik-Red'kov-1986}
    Bogush, A.A., Otchik, V.S., Red'kov, V.M.:
    Vector field in de Sitter space.
    Vesti AN NSSR. {\bf 1}, 58--62 (1986)


\bibitem{Mishima-Nakayama-1987}
Takashi Mishima, Akihiro Nakayama:
Particle production in de Sitter spacetime.
 Progr. Theor. Phys.  {\bf  77},  218--222 (1987)








\bibitem{Polarski-1989}
Polarski, D.:
The scalar wave equation on static de Sitter and
anti-de Sitter spaces.
 Class. Quantum Grav. {\bf 6},  893--900 (1989)


\bibitem{Suzuki-Takasugi-1996}
  Suzuki H.,  Takasugi, E.:
Absorption Probability of De Sitter Horizon for Massless Fields with Spin.
Mod. Phys. Lett. A   (1996) {\bf 11},  431--436;




\bibitem{Suzuki-Takasugi-Umetsu-1998}
 Suzuki, H.,  Takasugi,  E.,  Umetsu,  H.:
 Perturbations of Kerr-de Sitter Black Hole and Heun's Equations.
Prog. Theor. Phys.    {\bf 100},  491--505 (1998)


\bibitem{Suzuki-Takasugi-Umetsu-1999}
 Suzuki,H.,  Takasugi,  E.,  Umetsu,  H.:
 Analytic Solutions of Teukolsky Equation in Kerr-de Sitter and Kerr-Newman-de Sitter Geometries
 Prog. Theor. Phys.    {\bf 102},  253--272 (1999) [ arXiv:gr-qc/9905040]



\bibitem{Suzuki-Takasugi-Umetsu-2000}
  Suzuki, H.,  Takasugi,  E.,  Umetsu, H.:
 Absorption rate of the Kerr-de Sitter black hole and the Kerr-Newman-de Sitter black hole
Prog. Theor. Phys.   {\bf 103},  723--731 (2000) [arXiv:gr-qc/9911079]





\bibitem{De Broglie-1934(1)}
 De Broglie, L.:
 L'\'equation d'ondes du photon.
  C. R. Acad. Sci. Paris.  {\bf  199},  445--448 (1934)


\bibitem{De Broglie-1934(2)}
De Broglie, L.:
Une nouvelle conception de la lumi\`ere.  Paris,
(1934)



\bibitem{De Broglie-Winter-1934}
De Broglie, L.,  Winter M.J.:
 Sur le spin du photon.
  C. R. Acad. Sci. Paris.  {\bf 199},  813--816 (1934)




\bibitem{Petiau-1936}
Petiau, G.:
University of Paris thesis.
 Acad. Roy. de Belg. Classe Sci. Mem.   {\bf  2},  (1936)


\bibitem{Proca-1938}
Proca, A.:
 Th\'{e}orie  non relativiste  des particules \`{a} spin entier.
Journ. Phys. Rad.  {\bf  9},  61--66,  (1938)



\bibitem{Proca-1946}
Proca, A.:
 Sur les \'equations relativistes des particules \'el\'ementaires.
  C. R. Acad. Sci. Paris.  {\bf  223},   270--272 (1946)


\bibitem{Duffin-1938}
  Duffin, R.Y.:
  On the characteristic matrices of covariant systems.
 Phys. Rev.  {\bf  54},  1114--1114 (1938)


\bibitem{Kemmer-1939}
Kemmer, N.:   The particle aspect of meson theory.
Proc. Roy. Soc. London. A.  {\bf 173},  91--116  (1939)


\bibitem{Kemmer-1943}
Kemmer, N.:
 The algebra of meson matrices.
 Proc. Camb. Phil. Soc.  {\bf  39},  189--196 (1943)


\bibitem{Bhabha-1939}
 Bhabha,  H.J.:
  Classical theory of meson.
  Proc. Roy. Soc. London. A.  {\bf172},  384--409 (1939)




\bibitem{Belinfante-1939(1)}
  Belinfante, F.J.:
The undor equation of the meson field.
 Physica.  {\bf 6},  870--886 (1939)


\bibitem{Belinfante-1939(2)}
  Belinfante,  F.J.:
 On the spin angular momentum of meson.
Physica.  {\bf 6},  887--898  (1939)


\bibitem{Sakata-Taketani-1940}
Sakata, S.,   Taketani, M.:
 On the wave equation of the meson.
Proc. Phys. Math. Soc. Japan.  {\bf  22}, 757--770 (1940);
Reprinted  in:  Suppl. Progr. Theor. Phys.  {\bf 22},  84--97  (1955)


\bibitem{Tonnelat-1941}
  Tonnelat, M.A.:
 Sur la th\'{e}orie du photon dans un espace de
Riemann.  Ann. Phys. N.Y.  {\bf  15},  144 (1941)



\bibitem{Schrodinger-1943(1)}
Schr\"{o}dinger, E.:
Pentads, tetrads, and triads of meson matrices.
Proc. Roy. Irish. Acad. A.  {\bf  48},  135--146 (1943)


\bibitem{Schrodinger-1943(2)}
 Schr\"{o}dinger, E.:
 Systematics  of meson matrices.
 Proc. Roy. Irish. Acad.  {\bf  49},  29--42 (1943)




\bibitem{Hitler-1943}
  Hitler, W.:
 On the particle equation of the meson.
 Proc. Roy. Irish. Acad.  {\bf 49},  1 (1943)




\bibitem{Harish-Chandra-1946}
Harish-Chandra:
The correspondence between  the particle and wave
aspects of the meson and the photon.
  Proc.  Roy. Soc. London. A  {\bf  186},  502--525 (1946)


\bibitem{Harish-Chandra-1947(1)}
Harish-Chandra:
 On the algebra of the meson matrices.
  Proc. Camb. Phil. Soc.  {\bf  43}, 414--421  (1947)


\bibitem{Harish-Chandra-1947(2)}
Harish-Chandra:
On relativistic wave equation.
 Phys. Rev.  {\bf  71},   793--805 (1947)



\bibitem{Hoffmann-1947}
 Hoffmann, B.:
 The vector meson field and projective relativity.
 Phys. Rev.  {\bf  72},  458--465 (1947)



\bibitem{Utiyama-1947}
 Utiyama, R.:
 On  the interaction of mesons with the  gravitational field.
Progr. in Theor. Phys.  {\bf 2},  38--62 (1947)


\bibitem{Gel'fand-Yaglom-1948}
 Gel'fand, I.M.,   Yaglom, A.M.:
  General relativistic-invariant equations and
infinitedimemsional representations of the Lorentz group.
JETF.   {\bf  18},  703--733 (1948)



\bibitem{Schouten-1949}
 Schouten, J.A.:
On meson fields and conformal  transformations.
  Rev. Mod. Phys.    {\bf  21},  421--424 (1949)


\bibitem{Gupta-1950}
 Gupta, S.N.:
 Theory of  longitudinal photons in  quantum electrodynamics.
  Proc. Roy. Soc. London. A  {\bf 63},  681--691 (1950)




\bibitem{Bleuler-1950}
 Bleuler, K.:
Eine neue  Methode  zur  Behandlung  der
longitudinalen und skalaren Photonen.
 Helv. Phys. Acta.  {\bf  23},   567--586 (1950)


\bibitem{Fujiwara-1955}
 Fujiwara, I.:
 On the wave equation for spin 1 in Hamiltonian form.
 Progr. Theor. Phys.   {\bf 14},  166--167 (1955)






\bibitem{Borgardt-1956}
  Borgardt, A.A.:
 Matrix aspects of boson theory.
JETP.  {\bf 30},  334--341  (1956)





\bibitem{Borgardt-1958}
 Borgardt, A.A.:   Wave equation for a photon.
JETP.  {\bf 34},  1323--1325  (1958)






\bibitem{Kuohsien-1957}
Kuohsien, T.:
 Sur les theories matricielles du photon.
 C.  R.  Acad. Sci. Paris.  {\bf  245},  141--144 (1957)


\bibitem{Hjalmars-1961}
 Hjalmars, S.:
 Wave equations for scalar and vector particles in
gravitational fields.
 J. Math. Phys.  {\bf  2},  663--666 (1961)


\bibitem{Bogush-Fedorov-1962}
   Bogush, A.A.,  Fedorov, F.I.:   On properties of the Duffin-Kemmer
matrices.
Doklady AN BSSR.   {\bf   6},   81--85  (1962)



\bibitem{Beckers-Pirotte-1968}
Beckers, J.,  Pirotte, C.:
Vectorial meson equations in relation to
photon description.  Physica.  {\bf 39},   205 (1968)


\bibitem{Casanova-1969}
Casanova, G.:
Particules neutre de spin 1.
 C. R. Acad. Sci.  Paris, A.  {\bf 268}, 673--676 (1969)


\bibitem{Krivski-Romamenko-Fushchych-1969}
  Krivski, I.Yu.,  Romamenko, G.D.,   Fushchych, V.I.:
 Equation of the Duffin-Kemmerr type in 5-dimensional Minkowski space.
 Teor. Matem. Fiz.  {\bf 1},  242--250 (1969)


\bibitem{Goldman-Tsai-Yildiz-1972}
Goldman, T.,   Tsai, W.,   Yildiz, A.:
 Consistency of spin one theory.   Phys. Rev. D.  {\bf  5},  1926--1930 (1972)




\bibitem{Fushchych-Nikitin-1977}
 Fushchych,  W.I.,    Nikitin, A.G.:
 On the new invariance group of the
Dirac and  Kemmer-Duffin-Petiau equations.
 Lett. Nuovo Cim.   {\bf  19},  347--352 (1977)



















\bibitem{Lunardi-Pimentel-Teixeira-Valverde-2000}
 Lunardi, J.T.,  Pimentel, B.M.,  Teixeira, R.G.,   Valverde, J.S.:
 Remarques  on the Duffin-Kemmer-Petiau  theory and gauge invariance.
 Phys. Lett. A.   {\bf  268}, 165--173 (2000)


\bibitem{Fainberg-Pimentel-2000}
 Fainberg, V.Ya.,   Pimentel, B.M.:
 Duffin-Kemmer-Petiau and
Klein-Gordon-Fock Equations for Electromagnetic, Yang-Mills and
external Gravitational Field Interactions: proof of equivalence.
 Phys.Lett.  {\bf  271},  16--25 (2000)


\bibitem{De Montigny-Khanna-Santana-Santos-Vianna-2000}
De Montigny, M.,   Khanna, F.C.,    Santana, A.E.,    Santos, E.S.,
Vianna J.D.M.:
  Galilean covariance and and the Duffin-Kemmer-Petiau
equation.
 J. Phys. A.  {\bf   33},   273--278 (2000)









\bibitem{Lunardi-Pimentel-Valverde-Manzoni-2002}
 Lunardi, J.T.,    Pimentel, B.M.,  Valverde, J.S., Manzoni L.A.:
  Duffin-Kemmer-Petiau theory in the causal approach.
 Int. J. Mod. Phys. A. {\bf   17},   205--227 (2002)


\bibitem{Casana-Pimente-Lunardi-Teixeira-2002}
Casana, R.,   Pimentel, B.M.,   Lunardi, J.T.,  Teixeira, R.G.:
 Free electromagnetic field in Riemannian space-time via DKP theory.
Int. J. Mod.  Phys. A. {\bf 17},   4197--4202  (2002)

\bibitem{Casana-Fainberg-Pimentel-Lunardi-Teixeira-2003}
  Casana, R.,  Fainberg, V.Ya.,    Pimentel,  B.M.,    Lunardi, J.T.,     Teixeira R.G.:
Massless DKP fields in Riemann-Cartan space-times.
  Class.  Quantum Grav.  {\bf  20},  No 11,  2457--2465 (2003)











\bibitem{Red'kov-1998(1)}
 Red'kov, V.M.:
 Generally relativistical Daffin-Kemmer formalism
and behaviour of quantum-mechanical particle of spin $1$
in the Abelian monopole field.     arxiv:quant-ph/9812007

\bibitem{Red'kov-1998(2)}
  Red'kov V.M.:
On discrete symmetry for spin 1/2 and spin 1 particles in external
monopole field  and quantum-mechanical property of self-conjugacy.  arxiv:quant-ph/ 9812066





\bibitem{Bogush-Kisel-Tokarevskaya-Red'kov-2002(1)}
 Bogush, A.A.,   Kisel, V.V.,  Tokarevskaya, N.G.,   Red'kov, V.M.:
  Non-relativistic approximation in generally covariant
    of a vector particle.  Vesti NANB, Ser. fiz.-mat. {\bf  2},  61--66  (2002).

\bibitem{Bogush-Kisel-Tokarevskaya-Red'kov-2002(2)}
 Bogush, A.A.,   Kisel, V.V.,  Tokarevskaya, N.G.,   Red'kov, V.M.:
Duffin-Kemmer-Petiau formalism reexamined: non-relativistic
approximation for spin 0 and spin 1 particles in a Riemannian
space-time. Annales de la Fondation Louis de Broglie.  {\bf  32}, 355--381 (2007);
arXiv:0710.1423v1 [hep-th] 7 Oct 2007.





\bibitem{Schrodinger-1938}
 Schr\"{o}dinger, E.:
 The ambiguity of the wave function.
 Annalen der Physik. {\bf  32},  49--55 (1938)





\bibitem{Pauli-1939}
 Pauli, W.:
   \"{U}ber die Kriterium f\"{u}r  Ein-oder Zweiwertigkeit  der
Eigenfunktionen in der Wellenmechanik.
  Helv. Phys. Acta.  1939. {\bf  12}, 147--168 (1939)


\bibitem{Newman-1961}
  Newman, E.T.:
New Approach to Einstein and Maxwell-Einstein field equations.
J. Math. Phys.  {\bf  2},   674--676 (1961)


\bibitem{Wigner-1927}
Wigner,E.:
 Einige Folgerungen aus der Schr\"{0}dingerschen
Theorie f\"{u}r die Termstrukturen
   [Some consequences from Sch\"{o}"dinger's theory for term structures],
    Zeitschrift f\"{u}"r Physik. {bf 43},  601--623 (1927).
    Reprinted in: L. C. Biedenharn and H. van Dam,
    Quantum Theory of Angular Momentum, Academic Press, New York (1965)


\bibitem{Goldberg-Macfarlane-Newman-Rohrlich-Sudarshan-1967}
 Goldberg,  J.N.,    Macfarlane, A.J.,    Newman, E.T.,   Rohrlich, F.,
  Sudarshan E.C.G.:
Spin-s spherical harmonics and $\partial \hspace{-2.5mm}/ $.
 J. Math. Phys.  {\bf  8}, 2155--2161 (1967)





\bibitem{Dray-1985}
 Dray,  T.J.:
The relationship between monopole harmonics and spin-weighted
spherical harmonics.  J. Math. Phys. {\bf  26}, 1030--1033 (1985)


\bibitem{Dray-1986}
 Dray, T.J.:
A unified  treatment of Wigner D functions,
spin-weighted spherical harmonics, and monopole harmonics.
  J. Math. Phys. {\bf  27}, 781--792 (1986)



\bibitem{Krolikowski-Turski-1986}
 Krolikowski, W.,  Turski,  A.:
 Relativistic two-body equation for one Dirac and one Duffin-Kemmer-Petiau particle,
 consistent with the hole theory.
 Acta Phys. Polon.  B. {\bf  17},  75--81 (1986)

\bibitem{Turski-1986}
 Turski, A.:
Method of separating the  angular  coordinates in
two-body wave equation with spin.
 Acta Phys. Polon. B. {\bf  17},  337--346 (1986)


\bibitem{Red'kov-1998(3)}
    Red'kov, V.M.:
    Generally relativistical Tetrode -- Weyl -- Fock -- Ivanenko formalism
    and behavior of quantum-mechanical particles of spin $1/2$
    in the Abelian monopole field.  arXiv.org/quant-ph/9812002



\bibitem{Red'kov-1998(4)}
   Red'kov, V.M.:
   Generally relativistical Daffin-Kemmer formalism
   and behavior of quantum-mechanical particle of spin $1$
   in the Abelian monopole field.  arXiv.org/abs/quant-ph/9812007

\bibitem{Red'kov-1999(1)}
   Red'kov, V.M.:
   The doublet of Dirac fermions in the field of the non-Abelian
   monopole, isotopic chiral symmetry, and parity selection rules. arXiv.org/abs/quant-ph/9901011.




\bibitem{Red'kov-1999(2)}
   Red'kov,  V.M.:
   On intrinsic structure of wave functions of
   fermion triplet in external monopole field. arXiv.org/abs/quant-ph/9902034.















\bibitem{Varshalovich-Moskalev-Hersonskiy-1975}
 Varshalovich, D.A.,  Moskalev, A.N.,  Hersonskiy, V.K.:
Quantum theory of angular moment. Nauka, Leningrad,   (1975)






\bibitem{Weinberg-1972}
 Weinberg, S.:
Gravitation and Cosmology: Principles and
Applications of the General Theory of Relativity. John Wiley \&
Sons. Inc., New York, (1972)




\bibitem{Hawking-Ellis-1973}
Hawking, S.W., Ellis, G.F.R.:
 The large scale structure of spacetime.
Cambridge University Press, Cambridge (1973)

\bibitem{Birrel-Davies-1982}
N.D. Birrel,  P.C.W. Davies.
Quantum fields in curved space. Cambridge
University Press (1982)

\bibitem{Chandrasekhar-1983}
S. Chandrasekhar, The Mathematical Theory of Black
Holes, Oxford University Press, Oxford, (1983)


 \bibitem{Penrose-Rindler-1984}
Penrose R., Rindler W.
 Spinors and space-time. Volume I: Two-spinor calculus and relativistic
fields. Cambridge University Press (1984)




\bibitem{Lorentz-1904}
 Lorentz, H.: Proc. Electromagnetic phenomena in a system
moving with any velocity less than that of light Royal Acad. Amsterdam.  {\bf 6},  809--831  (1904)

\bibitem{Poincare-1905}
  Poincar\'e,  H.:
   Sur la dynamique de l'\'{e}lectron.
   C. R. Acad. Sci. Paris.   140,  1504 - 1508 (1905);
 Sur la dynamique de
l'\'{e}lectron.    Rendiconti del Circolo Matematico di Palermo.  {\bf 21},
129--175 (1906)





\bibitem{Einstein-1905}
  Einstein, A.:
   Zur Elektrodynamik der bewegten K\"orper.
     Annalen der Physik.  {\bf 17},  891--921 (1905)

\bibitem{Minkowski-1908}
  Minkowski H.  Nachrichten von der K\"{o}niglichen Gesellschaft der Wissenschaften zu G\"{o}ttingen,
    mathematisch-physikalische Klasse.  {\bf 53},  111 (1908);  Math. Ann. {\bf 68},  472--525 (1910)

\bibitem{Silberstein-1907}
  Silberstein, L:.   Elektromagnetische Grundgleichungen in
bivectorieller Behandlung. Ann. Phys.  {\bf 22},  579--586 (1907)

\bibitem{Weber-1901}
  Weber, H.:
   Die partiellen Differential-Gleichungen der mathematischen Physik nach Riemann's Vorlesungen.
    Friedrich Vieweg und Sohn. Braunschweig.   348 (1901)

\bibitem{Marcolongo-1912}
  Marcolongo, R.
  Les transformations de Lorentz et les \'equations
de l'\'electrodynamique.
 Annales de la Facult\'{e} des Sciences de Toulouse.  {\bf 4}, 429--468 (1912)

\bibitem{Bateman-1915}
  Bateman, H.:
   The Mathematical analysis of electrical and optical wave-motion on the basis
   of Maxwell's equations. Cambridge University Press (1915)

\bibitem{Oppenheimer-1931}
  Oppenheimer, J.:
   Note on light Quanta and the electromagnetic
field. Phys. Rev. {\bf 38},  725--746 (1931)

\bibitem{Majorana-1931}
 Majorana E. Scientific Papers.
 Unpublished. Deposited at the "Domus Galileana". Pisa, quaderno 2, p. 101/1; 3, p. 11, 160; 15, p. 16; 17, p. 83, 159.



\bibitem{Bialynicki-Birula-1994}
 Bialynicki-Birula, I.:
 On the wave function of the photon. Acta Phys. Polon.  86,  97 - 116 (1994);
 Photon wave function.
   Progress in Optics. {\bf 36}, 248--294 (1996)



\bibitem{Sipe-1995}
 Sipe, J.:
    Photon wave functions.  Phys. Rev. A  {\bf 52},  1875--1883 (1995)

\bibitem{Gersten-1997}
Gersten, A.:
 Maxwell equations as the one-photon quantum equation
 Found. of Phys. Lett.  {\bf 12}, 291--298 (1998) [arXiv:quant-ph/9911049]

\bibitem{Esposito-1997}
 Esposito, S.:
  Covariant Majorana formulation of
electrodynamics. Found. Phys.{\bf  28}, 231--244  (1998) [arXiv:hep-th/9704144]

\bibitem{Dvoeglazov-1998}
 Dvoeglazov, V.:
 Historical note on relativistic theories of
electromagnetism. Apeiron.  5,  69 - 88 (1998) [arXiv:physics/9802039]

\bibitem{Ivezic-2006}
 Ivezic, T.:   Lorentz invariant Majorana formulation of the
field equations and Dirac-like equation for the free photon.  Electronic J. Theor. Phys.  {\bf 3}, 131--142 (2006)

\bibitem{Varlamov-2003}
 Varlamov, V.:
 About algebraic foundations of Majorana -- Oppenheimer
quantum electrodynamics and de Broglie -- Jordan neutrino theory
of light.
 Ann. Fond. L. de Broglie.  {\bf 27}, 273--286 (2003)


\bibitem{Red'kov-Bogush-Tokarevskaya-Spix-2009}
 Red'kov, V.M.,   Bogush A.A.,  Tokarevskaya N.G., George J. Spix.
Majorana-Oppengeimer approach to Maxwell electrodynamics in Riemannian space-time.
 Proc.  of 14th International School \& Conference "Foundation \& Advances
in Nonlinear Science",
September 22-25,  2008, Minsk. Pages 20-49 [arxiv.org/0905.0261]


\bibitem{Bogush-Krylov-Ovsiyuk-Red'kov-2009}
 Bogush,  A.A.,  Krylov, G.G.,  Ovsiyuk, E.M.,   Red'kov, V.M.:
 Maxwell equations in complex form of Majorana-Oppenheimer,
  solutions with cylindric symmetry in Riemann $S_{3}$  and Lobachevsky $H_{3}$ spaces.
      Ricerche di matematica. 2009.  DOI 10.1007/s11587-009-0067-8


\bibitem{Book-2009}
 Red'kov, V.M.:
Fields in Riemannian space  and the Lorentz group.
Belorussian Science, Minsk, (2009)













\end{thebibliography}
\end{document}